\documentclass[a4paper,11pt]{article}
\usepackage[nodayofweek]{datetime}
\usepackage{amsmath}
\usepackage{amsfonts}
\usepackage{algorithm,algorithmic}
\usepackage{thmtools,thm-restate}
\usepackage{enumitem} 
\usepackage{float}
\usepackage[margin=2.5cm]{geometry}
\usepackage{graphicx}
\usepackage[labelfont=bf,labelsep=period]{caption}
\usepackage{times}
\usepackage{url}
\usepackage{xspace}
  
\usepackage{amsmath}
\allowdisplaybreaks[2]

\usepackage{amssymb}
\usepackage[export]{adjustbox}

\usepackage{amsthm} 
\newtheoremstyle{plain-boldhead}
  {\topsep}
  {\topsep}
  {\itshape}
  {}
  {\bfseries}
  {.}
  { }
  {\thmname{#1}\thmnumber{ #2}\thmnote{ (\bfseries #3)}}
\newtheoremstyle{definition-boldhead}
  {\topsep}
  {\topsep}
  {\normalfont}
  {}
  {\bfseries}
  {.}
  { }
  {\thmname{#1}\thmnumber{ #2}\thmnote{ (\bfseries #3)}}
\theoremstyle{plain-boldhead}   
\newtheorem{theorem}{Theorem}

\newtheorem{lemma}[theorem]{Lemma}
\newtheorem{corollary}[theorem]{Corollary}

\theoremstyle{definition-boldhead}
\newtheorem{definition}{Definition}
\newtheorem{remark}{Remark}

\floatstyle{ruled}
\newfloat{algo}{htbp}{algo}
\floatname{algo}{Algorithm}

\newcommand{\E}{\mathrm{E}}

\newcommand{\var}[1]{\textit{#1}}
\newcommand{\op}[1]{\textsl{#1}}

\newcommand{\true}{\textsc{true}\xspace}

\newcommand{\CB}{\ensuremath{\mathcal{B}}\xspace}

\newcommand{\CM}{\ensuremath{\mathcal{M}}\xspace}

\newcommand{\CP}{\ensuremath{\mathcal{P}}\xspace}

\newcommand{\CS}{\ensuremath{\mathcal{S}}\xspace}
\newcommand{\CT}{\ensuremath{\mathcal{T}}\xspace}

\def \ifempty#1{\def\temp{#1} \ifx\temp\empty }

\newdateformat{simple}{\THEDAY\ \monthname[\THEMONTH]\ \THEYEAR}
\simple

\begin{document}

\title{\bf Generalizing Weighted Trees: A Bridge from Bitcoin to GHOST}

\author{Ignacio Amores-Sesar \\ 
  University of Bern \\
  \url{ignacio.amores@inf.unibe.ch}
  \and Christian Cachin\\
  University of Bern\\
  \url{cachin@inf.unibe.ch}
  \and Anna Parker\footnote{Work done at the University of Bern.}\\
  \url{anya.m.parker@gmail.com}
}

\date{\today}

\maketitle

\begin{abstract}\noindent
  Despite the tremendous interest in cryptocurrencies like Bitcoin and
  Ethereum today, many aspects of the underlying consensus protocols are
  poorly understood.  Therefore, the search for protocols that improve
  either throughput or security (or both) continues.  Bitcoin always
  selects the longest chain (i.e., the one with most work).  Forks may
  occur when two miners extend the same block simultaneously, and the
  frequency of forks depends on how fast blocks are propagated in the
  network.  In the GHOST protocol, used by Ethereum, all blocks involved in
  the fork contribute to the security.  However, the greedy chain selection
  rule of GHOST does not consider the full information available in the
  block tree, which has led to some concerns about its security.

  This paper introduces a new family of protocols, called Medium, which takes the structure
  of the whole block tree into account, by weighting blocks differently
  according to their depths.  Bitcoin and GHOST result as special cases.
  This protocol leads to new insights about the security of Bitcoin and
  GHOST and paves the way for developing network- and application-specific
  protocols, in which the influence of forks on the chain-selection process
  can be controlled.  It is shown that almost all protocols in this family 
  achieve strictly greater throughput than Bitcoin (at
  the same security level) and resist attacks that can be mounted against
  GHOST.
\end{abstract}

\maketitle

\section{Introduction} 

Since Nakamoto revealed the Bitcoin protocol \cite{nakamoto2019bitcoin} as a blueprint for a decentralized payment system, many other protocols have been introduced with the goal of improving Bitcoin. 
The basic principle of these decentralized payment systems is that a distributed data structure, called the blockchain, is maintained by parties (also called \emph{miners}) that run a distributed protocol. Transactions are grouped into blocks, which are later added to the blockchain when specific properties have been fulfilled.
Most improvements to Bitcoin aim at processing more transactions and achieving higher throughput without degrading security because Bitcoin is severely limited in this sense~\cite{DBLP:conf/fc/CromanDEGJKMSSS16}. The GHOST protocol~\cite{DBLP:conf/fc/SompolinskyZ15}, for example, lets all mined blocks contribute to the security by considering subtrees of blocks, whereas Bitcoin relies only on the blocks in the longest chain. GHOST, however, does not take into consideration how the blocks are structured and counts \emph{all} blocks in a subtree in the same way. This introduces a potential vulnerability to consensus, which can be exploited by an adversary with strong influence over the network, as exemplified in a balance attack~\cite{DBLP:conf/dsn/NatoliG17}.

Whenever a miner produces a block, the miner places it in some position with respect to the previously produced blocks (by including their hashes in the new block). Thus, a protocol execution constructs a tree, in which every node is a block $B_i$ and an edge $(B_i,B_j)$ denotes that $B_i$ includes the hash of $B_j$. This tree can be used to understand the placement of newly mined blocks in Bitcoin and in GHOST within a common framework. The chain that the miners extend is called the \emph{main chain}. The key difference between such protocols lies the way how this main chain is selected.

As a miner in Bitcoin always selects the longest chain in the tree (technically, the one with most work, but we ignore this subtlety here) and extends this chain by one block. The security relies intuitively on the rule that only the longest chain grows, unless two parties mine concurrently and thereby create a fork. This may happen when a party mines without receiving the last block mined before. Forks limit the throughput of a network, and they typically occur more often when the block production rate increases compared to the message delay in the network.

On the other hand, GHOST determines the main chain by extracting more information from the tree. Starting from the genesis block, it iteratively selects the block with the heaviest subtree (defined by the number of blocks in the subtree of the block) until it reaches a leaf block. When a miner produces a new block, it appends this to the last block selected by this rule. The intuition is that also forked blocks (and their miners) contribute to the security of the blocks they point to. However, all blocks are counted in the  same way regardless of their position in the subtree.  This actually loses considerable information about the tree structure and may introduce vulnerabilities.

In this paper, we introduce the \emph{Medium protocol}\footnote{Medium, in occultism, a person reputedly able to make contact with the world of spirits, especially while in a state of trance~\cite{mediumbritannica2021}.} that takes into account the structure of the block tree in a way that generalizes both Bitcoin and GHOST. Medium computes a \emph{weight} for a subtree using a \emph{polynomial} in a \emph{weight coefficient}~$c$, which determines the influence of the tree structure on chain selection. This results in a family of Medium protocols, each one uniquely defined by some~$c$.

Specifically, we introduce a weight function
\begin{equation}
  \omega:\CB\times\CT\to\mathbb{R}_{> 0}
\end{equation}
for a block $B\in\CB$ in a tree $T\in\CT$,
defined by $\omega(B,T)=c^{d(B)}$, where $d(B)$ denotes the depth of $B$ in $T$ and $c\geq 1$. The selection rule of GHOST can be interpreted as the particular case of $c=1$ (up to the way of breaking ties for trees with equal weight), and Nakamoto consensus results in the limit for~$c\to\infty$. Thus, Medium generalizes GHOST and Bitcoin, so that they can be compared in a comprehensive way to all protocols in the Medium family.

The weight function intuitively takes up the idea behind GHOST that every block contributes to the security and combines it with Bitcoin's feature that deeper blocks are more relevant. Thus, also forked blocks influence the main chain selection process, but longer chains are still more desirable. 

The weight coefficient determines the extent to which forks contribute to main chain selection in relation to the contribution of chain length.

We show that Medium is secure against well-known attacks on GHOST. In particular, a balance attack always fails after a finite number of rounds. We show that protocols with larger weight coefficients are in general safer from attacks, but may have lower throughput. There is thus a continuum of weight coefficient values, leading to the ability to find a protocol with optimized throughput and safety, depending on the network and the user's requirements.

To analyze the security of Medium, we adopt the model of Kiayias and Panagiotakos~\cite{DBLP:conf/latincrypt/KiayiasP17}, which allows us to prove security against attacks on consensus, such as double spending~\cite{nakamoto2019bitcoin}, block withholding~\cite{DBLP:journals/cacm/EyalS18}, and eclipse~\cite{DBLP:conf/uss/HeilmanKZG15}.
Specifically, we prove that the Medium protocol family satisfies three main properties in a synchronous network.
Firstly, the \emph{weight and length} of the main chain \emph{increase} over time. This means that the protocol is live, adding ever more transactions to the blockchain, and also that the cost of reverting past transactions increases with time.
Secondly, the main chain \emph{contains} at least a \emph{fraction of honest blocks}, i.e., blocks not mined by the adversary. This ensures that transactions of honest parties are eventually added to the main chain and executed.
And lastly, the main chain of all the honest parties contains a \emph{common prefix} that increases over time. This means that once a transaction has been in the main chain for long enough, it remains in the main chain.
We use these properties to ultimately construct a decentralized payment system, where the blockchain is a robust public transaction ledger, following the notions of Kiayias \emph{et al.}~\cite{DBLP:conf/eurocrypt/GarayKL15,DBLP:conf/latincrypt/KiayiasP17}.

The results illustrate how Medium forms a bridge between Bitcoin and GHOST, allowing a deeper understanding of them; Medium can also improve other constructions that rely on Bitcoin or GHOST.

\subsection{Related work}

Garay \emph{et al.}'s Bitcoin Backbone~\cite{DBLP:conf/eurocrypt/GarayKL15} is the first in-depth formalization of the Bitcoin protocol and represents an important step for understanding the security of blockchains. They analyze the protocol in synchronous and in partially synchronous networks. Kiayias and Panagiotakos~\cite{DBLP:conf/latincrypt/KiayiasP17} expand the model and demonstrate the security of Bitcoin and GHOST against a variety of attacks.

In these security models, the adversary has only limited capability to prevent communication between honest parties. For instance, in the analysis of the eclipse attack~\cite{DBLP:conf/latincrypt/KiayiasP17}, the adversary may only control the communication between a fraction of the miners. More powerful attacks, however, could split the network in two and prevent any exchange between the parts. Such attacks threaten the security of Bitcoin and have even more severe consequences for GHOST. In particular, Natoli and Gramoli~\cite{DBLP:conf/dsn/NatoliG17} point out this issue under the name of a \emph{balance attack}.  Bagaria \emph{et al.}~\cite{DBLP:conf/ccs/BagariaKTFV19} show that such an attack on GHOST can perpetuate a fork indefinitely, leading to miners splitting their power between the two sides of the fork and the network never reaching consensus. The difference between these attacks is that Bagaria \emph{et al.}~\cite{DBLP:conf/ccs/BagariaKTFV19} assume the adversary has the ability to partition the network for a given amount of time. It is exactly such an attack that we aim to prevent by choosing a proper weight coefficient.

We note that Kiayias and Panagiotakos~\cite{DBLP:conf/latincrypt/KiayiasP17} present a unified description and security analysis of the GHOST and Bitcoin protocols. This analysis relies on a using a weight norm, however, and their analysis only holds for linear weight functions. For blockchains this means the weight of a subtree must increase linearly in relation to the number of blocks. This condition limits their analysis to boundary cases (e.g., Bitcoin and GHOST); it cannot be applied to Medium's polynomial weight functions. We present a different approach, which adopts much of their notation and builds on their methodology and models. This should facilitate comparison of the two protocols including the spectrum between them.

The existence of protocols achieving a higher throughput than both Bitcoin and GHOST is a well-known fact. Some of the most prominent examples are: BitcoinNG~\cite{DBLP:conf/nsdi/BitcoinNG}, Conflux~\cite{DBLP:journals/corr/Conflux}, and Prism~\cite{DBLP:conf/ccs/BagariaKTFV19}. The reason for studying the spectrum between GHOST and Bitcoin is that the previously mentioned protocols use either Bitcoin or GHOST as a building block. Hence, given that Medium has either better security or better throughput than Bitcoin or GHOST, these sophisticated protocols may inherit Medium's properties.

BitcoinNG~\cite{DBLP:conf/nsdi/BitcoinNG} uses Bitcoin's rule to elect leaders. These leader have then the ability to generate many blocks. However, the security of the protocol depends completely on these leader-election blocks.  Hence, a different rule for leader election at a higher security level, or with a higher ratio of leaders per unit of time, translates in an immediate upgrade of this protocol.

The main innovation behind Conflux~\cite{DBLP:journals/corr/Conflux} is its ability to include orphan blocks in the ledger. Conflux uses the GHOST's rule to agree on a main chain. Consequently, Conflux uses a secondary set of references in order to topologically order the complete DAG. Conflux then purifies this DAG to eliminate all the possible double-spendings and builds the ledger. Once again, a better rule for the selection of the main chain improves the totality of the protocol.

With regard to Prism~\cite{DBLP:conf/ccs/BagariaKTFV19}, finally, the situation is slightly more complex because its selection rule is more sophisticated. In this protocol, a block is not classified as valid or invalid depending on the value of its header. Instead, it is classified in several groups depending on the value of the hash function. One of these groups is invalid, another one allows the block to contribute with its transactions, but not to the chain selection, and another group contributes only to this chain selection. The security of this protocol relies exclusively in this last group, the chain selection inside this group follows a variation of Bitcoin. Hence, Medium's chain-selection rule could again be exploited to upgrade Prism.

\section{Model}
\label{sec:preliminaries}

\subsection{General definitions}

Similarly to the Bitcoin Backbone protocol~\cite{DBLP:conf/eurocrypt/GarayKL15}, the execution of the protocol takes place in rounds. At the start of each round, parties receive the messages sent to them in the previous round, then the parties perform specific operations and finish the round by specifying the messages they want to broadcast. 

A \emph{block} is defined as a tuple of the form $B = [s, x, i, ctr]$ with 
$s \in \{ 0, 1, \}^{\kappa}, x \in \{ 0, 1\}^*$, $ctr \in \mathbb{N}$ and $i \in \{1, ..., n\}$ (where $n$ is the total number of parties). Two cryptographic hash functions $G(\cdot)$ and $H(\cdot)$, which are modeled as random oracle functionalities~\cite{DBLP:conf/ccs/BellareR93} are used to define the validity of a block.

A block, mined by party $P_i$ is defined as \emph{valid} if it satisfies the condition
\[ (H(ctr, G(s, x,i)) < D) \wedge (ctr \leq q),\]
where $D$ is the difficulty parameter and $q$ is the maximum number of hash queries in a round. 

A \emph{chain} $C$ is a sequence of valid blocks, starting from the root block ($genesis(C)$) and extending to a final, head block ($head(C)$). For a chain to be valid each block in the chain must be valid and fulfill the condition that if a block $B = [s, x, i, ctr]$ extends block $B' = [s', x', i', ctr']$ in the chain, then $s = H(ctr', G(s', x', i'))$.  

We say a new block has been \emph{mined} if a valid block can be found that extends a chain in this valid manner. Since the difficulty parameter is $D$, the success probability of a single hashing query is $p = \frac{D}{2^\kappa}$, where $\kappa$ is the length of the hash. 

Miners that attempt to mine on the blockchain are referred to as \emph{parties}.

There are a total of $n$ mining parties, of these the adversary controls a maximum of $t$, the parties controlled by the adversary are called \emph{corrupted}. Parties running the protocol are called \emph{honest} and only communicate at the end of a round. When an honest party mines a block this block is referred to as an \emph{honest block}. When an corrupted party, controlled by the adversary, mines a block it is referred to as a \emph{corrupted block}. 

A round is called \emph{successful} if an honest party mines a block in that round, and \emph{uniquely successful} if only one honest party mines in that round.

The length of the chain $C$ is denoted by $\ell(C)$. When looking at a chain $C$, we say $C$ \emph{extends} another chain $C'$ if $C'$ is a prefix of $C$, we can then write $C' \preceq C$. The \emph{depth} of a block $B$ in the blockchain is the length of the path from that block to the genesis block. The tree of blocks mined by these parties is called the \emph{block tree}, each party has a \emph{local view} of the block tree which is comprised of all the valid mined blocks that it knows about. A \emph{fork} in the tree occurs when two parties mine on the same block, extending the same chain. Which chain and therefore which block in the tree is mined on is decided by each party according to the protocol, this chain is referred to as the \emph{main chain} and is chosen by the main chain selection algorithm. How the protocol handles forks must be defined in such an algorithm.

\subsection{Communication and mining}

We base our security analysis on the model used in the Bitcoin Backbone paper \cite{DBLP:conf/eurocrypt/GarayKL15}.
We assume there to be a set of $n$ parties, $\CP=\{P_1,...,P_n\}$ running the protocol, modeled as interactive Turing machines (ITM). An interactive Turing machine is a Turing machine with an input and an output tape that allow the Turing machines to communicate with other Turing machines and make decisions depending on the content of their input tape. The adversary is modeled as another ITM that corrupts up to $t$ parties at the beginning of the execution. These corrupted parties obey the adversary, in other words they may diverge from the normal execution of the protocol.  All the parties running the protocol and the adversary have access to two functionalities.

A \emph{diffusion functionality} implements communication among the parties, which is structured into \emph{synchronous} rounds. The functionality keeps a $\op{RECEIVE}_i$ string for each party $P_i$ and makes it available to $P_i$ at the start of every round. When a party $P_i$ instructs the diffusion functionality to $\op{BROASCAST}$ a message, $P_i$ is tagged as finished for this round. The adversary is allowed to read the string of any party at any moment during the execution and to see any messages broadcast by honest parties immediately. Furthermore, the adversary has a special message to indicate when it has finished sending its communications for a round and can write messages directly and selectively into $\op{RECEIVE}_i$ for any $P_i$. When all honest parties have finished the round, the diffusion functionality takes all messages that were broadcast by honest parties in the round and adds them to $\op{RECEIVE}_i$ for all parties. This models a \emph{rushing} adversary.

Every honest party communicates changes to its local view at the end of each round.  If an honest party finds a block in round~$r$, the new block is be received by all parties by the end of that round. Furthermore, even if the adversary causes a block to be received selectively by only some honest parties in round $r$, the block is seen by all honest parties at the end of round $r+1$.

The \emph{random oracle} is a functionality that can be queried in two different ways. If queried with input $x$ as calculation, the random oracle returns a random string of a given length $\kappa$ if it was not queried with $x$ before. If was previously queried with input $x$ it returns the same output as before. However, it can also be queried as verification with inputs $(x,y)$, the random oracle outputs 1 if it was queried, for calculation, before with input $x$ and the corresponding output was $y$. Otherwise it outputs 0.  (The separate verification queries let this differ from the standard random-oracle model, but this is necessary in our context~\cite{DBLP:conf/eurocrypt/GarayKL15}.) 

Any party has access to $q$ queries of the random oracle for calculation, the adversary has $q$ queries per corrupted party. The number of queries for verification is unbounded for honest parties, however the adversary has no access to verification queries. This has been called the \emph{$q$-bounded flat model}~\cite{DBLP:conf/eurocrypt/GarayKL15}.

\section{The Medium protocol}
\label{sec:Medium}

The \emph{Medium} protocol proceeds roughly like the Bitcoin and GHOST protocols~\cite{DBLP:conf/latincrypt/KiayiasP17} by arranging the received blocks into a tree, as also formalized by the Bitcoin Backbone protocol~\cite{DBLP:conf/eurocrypt/GarayKL15}. Bitcoin then selects the longest branch in the tree as its main chain, and GHOST constructs its main chain by greedily selecting the block with the heaviest subtree by number of blocks.
In Medium, the main chain is determined by always following the heaviest \emph{weighted} subtree, using the Medium weight function introduced here.

\begin{definition}[Weight]
The \emph{weight of a block} $B$ in a tree $T$ is given by
\[
  \omega_{c}(B,T)=c^{d(B)},
\]
where $d(B)$ denotes the depth of $B$ in $T$ when the Medium protocol is
instantiated with weight coefficient~$c$. 

\end{definition}

\begin{definition}[Tree Weight]
  The \emph{weight of a tree} $T$ is the sum of the weights of all 
  blocks of~$T$,
  \[
    \omega_c( T) = \sum_{B^\prime\in T}\omega_c(B^\prime, T).
  \]
\end{definition}

Notice that the contribution of each block to the tree weight
depends on the position of the block in the tree.  We define $T(B)$ to be
the subtree rooted at a block~$B$ and refer to the weight of $T(B)$ as the
\emph{tree weight}~$B$.

\subsection{Protocol details}

In more detail, each party starts a round with a local view of the block tree and its current main chain~$C$. To determine the new main chain, the protocol recursively iterates over the block tree, starting from the genesis block. At each block, the protocol extends the main chain with the child that has the heaviest tree weight, that is, by choosing the (polynomially weighted) heaviest subtree. Ties are broken by choosing the root of the subtree that results in the longest main chain, or if this would be the same, then by selecting the block that has been received earlier.  Extending the main chain through proof-of-work (\op{POW}) occurs similarly to the Bitcoin Backbone protocol.

\begin{algo}
  \small
  \begin{tabbing}
    xxxx\=xxxx\=xxxx\=xxxx\=xxxx\=xxxx\=xxxx\kill
    \textbf{function} $\op{Medium}(T,\omega_c)$
    \> \` //  a tree $T$ and a weight function $\omega_c$\\
    \>$B \leftarrow \op{root}(T)$\\
    \> \textbf{if} $\op{desc}(B) = \emptyset$ \textbf{then} \\
    \>\> \textbf{return} B \\
    \> \textbf{else}
    \` // break ties by larger depth of trees\\
    \>\> $B \leftarrow \op{argmax}\{ \omega_c(T(B')): B' \in \op{desc}(B)\}$ \\
    \>\> \textbf{return} $B \| \op{Medium}(T(B), \omega_c)$ \`// concatenate blocks \\
    \\
    \textbf{function} \op{$\omega_c(T)$}\` // weight function $\omega_c$ with coefficient $c$ \\
    \>$B \leftarrow \op{root}(T)$\\
    \> $\text{sum} \leftarrow 0$ \\
    \> \textbf{for} $B' \in \op{desc}(B)$ \textbf{do} \\
    \> \> $\text{sum} \leftarrow \text{sum} + \omega_c(T(B'))$ \\
    \textbf{return} $c \cdot \text{sum} + 1$ 
  \end{tabbing}
  \caption{Main chain selection algorithm}
  \label{algo:chain}

\end{algo}

The miner starts the round and checks the input string $\op{RECEIVE}_i$ for new blocks. The miner then runs $\op{update}()$ to extend its local tree and validate any received blocks. Then it runs the $\op{Medium}()$ algorithm, as illustrated in Algorithm \ref{algo:chain}, to determine its main chain.  If $\op{update}()$ has added a new block to the local tree, the miner broadcasts this new block again at the end of the round. 

After this is completed the miner can start running the $\op{POW}$ algorithm to try to mine a new block that can extend the main chain and fulfill the needed properties for validity. If the party mines such a block it uses the diffusion functionality to send a message with the block information to all parties at the end of the round, we call this \emph{broadcasting} the block. By broadcasting the blocks the party has accepted during a round again at the end of the round the protocol ensures other honest parties also receive the same blocks and can update their own trees accordingly. This ensures that if an adversary broadcasts in round $r$ to an honest party by the end of round $r+1$ all other parties also receive the block.  A formal description is included in Appendix~\ref{app:protocols}.

\subsection{Choice of the weight coefficient}
\label{ssec:weightcoefficient}

To make it harder for the adversary to perpetrate the balance attack. we may choose weight coefficients $c >1$ of a particular shape. Given a tree $T$, we can express its tree weight $\omega_c(T)$ as a polynomial in $c$ of degree $\ell$, 
\[ \omega_c(T) = a_0 c^0 + a_1 c^1 + ... + a_{\ell} c^\ell ,\]
where $\ell$ is the depth of the tree and the coefficient $a_i$ expresses how many blocks there are at level $i$ in the tree. We observe that $a_i \geq 1$ for $i \in \{0, .. ., \ell \}$; furthermore the total number of blocks in the tree is $N = \sum_i a_i$. We can use these polynomials to compare the weight of two different trees, $T_1$ and $T_2$. Two trees have equal weight whenever
\begin{align*} 0&= \omega_c(T_1)- \omega_c(T_2) \\ &= (a_{0,1} - a_{0,2})+ ... + (a_{max\{\ell_1, \ell_2\},1} - a_{max\{\ell_1, \ell_2\},2}) c^{max\{\ell_1, \ell_2\}}\end{align*}

Clearly, the weight of the two trees is the same if $c$ is a root of the polynomial resulting from their difference. If we want two trees of given depth $\leq\ell$ to have the same weight if and only if they have the same structure, we need to consider a weight coefficient $c$ that it is not a root of any polynomial of degree $\ell$ or less.

Consider the polynomial $f_{n,p}(X)=X^n-p$ with $p$ a prime number and $n\geq 1$, by Eisenstein's criteria~\cite{EisensteinCriteria}, this polynomial is irreducible on $\mathbb{Z}$. We define the set 
\begin{equation}
\CS_\ell=\{c: f_{n,p}(c)=0|c\in\mathbb{R}, n\geq \ell, p\ \text{prime}\},
\end{equation}
any constant taken from this set is  a root of an irreducible polynomial of degree at least $\ell$. Hence, to make sure that two trees of depth $\leq\ell$ have the same weight if and only if they have the same structure, it is enough to consider any element from~$\CS_\ell$.

\subsection{Relation with Bitcoin and GHOST}

\begin{figure}[t!]
  \centering
  \includegraphics[width=0.60\linewidth]{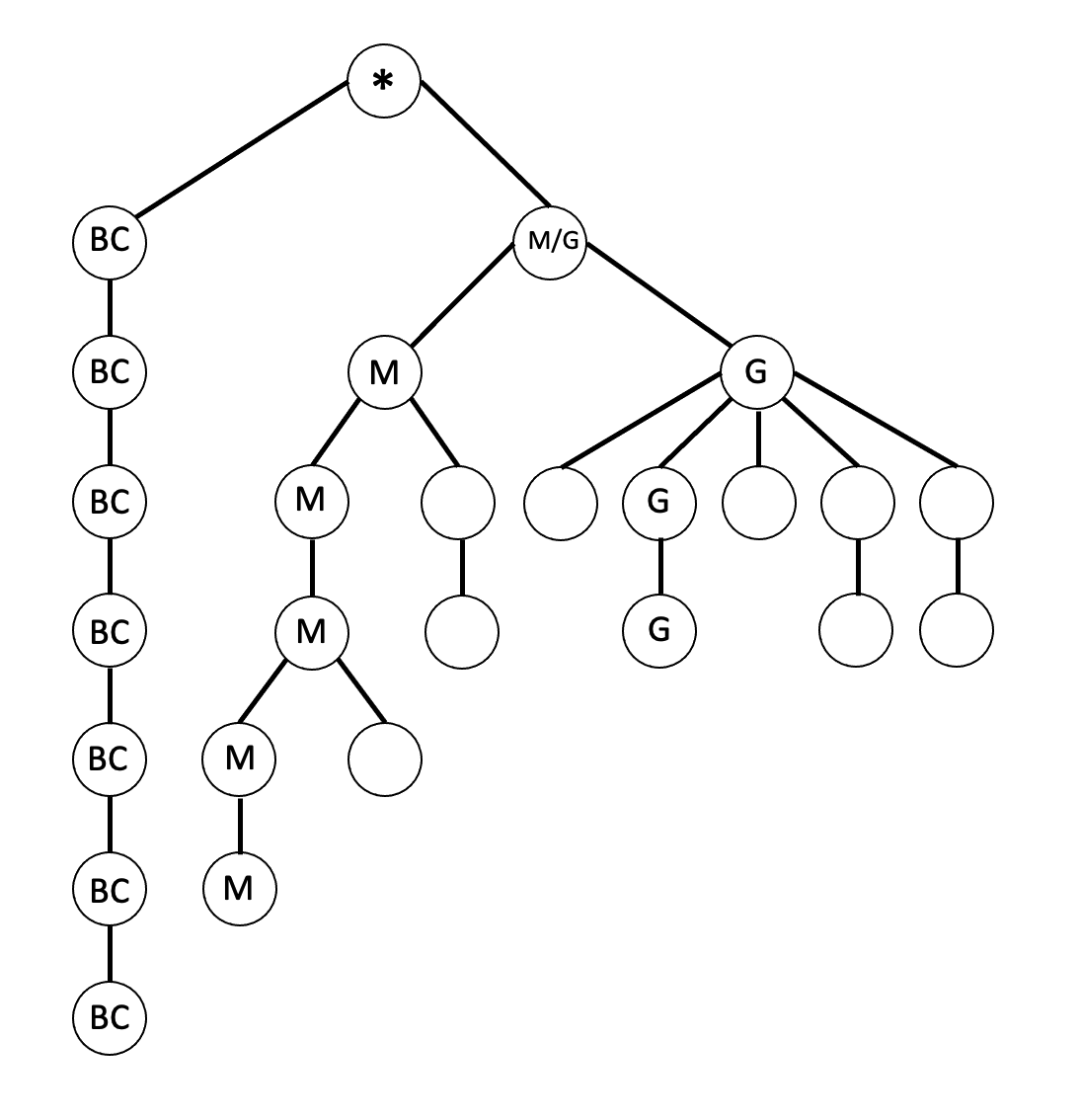}
  \vspace*{-3mm}
  \caption{In this example, where every blocks represents an equal amount of hashing work, different chains are selected by Bitcoin (BC), GHOST (G), and Medium (M) with $c=2$. Bitcoin simply selects the longest chain, but much less hashing power may have gone into this than into the other subtree. GHOST, however, selects the blocks in the larger subtree. One drawback of GHOST is the big loss of information about the structure of the ignored subtrees. Medium selects a chain that represents more hashing power, than the chain chosen by Bitcoin; at the same time, more structural information about the tree is taken into account by Medium than by GHOST.}
  \label{fig:subtrees}
\end{figure}

Above we explained how to select the weight coefficient to guarantee that trees of some bounded depth have the same weight if and only if they have the same structure. However, there are different choices of $c$ that are interesting to study. 

If we select $c=1$, our protocol reduces to the GHOST protocol. Additionally, the polynomial associated to the tree structure reduces to the number of blocks. In other words, we lose a huge amount of information regarding tree structure.

In the other extreme, if we consider increasing values of $c$,
the weight of a block in the tree is the same as the weight of $c$ blocks in the previous level. This difference increases with $c$, thus, when $c$ is large, we need a large number of blocks in the previous level to match the weight of a single block. This shows, intuitively, that the Medium protocol behaves like Bitcoin for $c\to\infty$ because the longest path in a subtree dominates its weight.

An execution that illustrates differences between Bitcoin, GHOST, and
Medium is shown in Figure~\ref{fig:subtrees}.

\section{Security analysis}\label{Security Analysis of Poly GHOST Backbone protocol}

The aim of this security analysis is to show that Medium is a robust transaction ledger, in other words, Medium satisfies liveness and persistence. To do this, we shall show that if a block is in the main chain and a sufficient number of blocks have been mined on this main chain after that block, so that these subsequent blocks weigh a predefined amount, then that block is stable (see Definition~\ref{def:stable} later). This means, the block remains in the main chain of any honest party except with negligible probability. We also show that in sufficiently many consecutive rounds there is always one honest block that enters the main chain and becomes stable.

We show this by establishing that the weight of the block tree increases in a specific manner during the execution of the protocol. This is done with the help of a \emph{typical execution}. This denotes an execution in which for any set of enough consecutive rounds, the random variables do not diverge form the expected value in a significant quantity. An execution is not typical with negligible probability. We also analyze how the tree produced by running the Medium protocol behaves, which permits us to specify the corresponding increase in weight. We determine upper and lower bounds for this weight increase, which hold except with negligible probability. We use these bounds to establish our version of the common prefix property. If we remove blocks according to a specific weight condition from the main chains of two honest parties, the resulting chains are a prefix of each other. Furthermore, if we remove blocks according to this weight condition for one honest party at round $r$, this chain is a prefix of the main chain of any honest party in all later rounds. By determining the minimal number of rounds needed to let the block tree grow by a specific amount, we can also show how the implied main chain becomes stable. With this knowledge, we will finally show that a minimal number of honest blocks are  produced in every consecutive subset of these rounds, that they are in the main chain, and that they remain stable.

\begin{table}
    \centering
    \small
    \begin{tabular}{|c|l|} 
    \hline
    \multicolumn{2}{| c |}{\textbf{Overview of Parameters and Variables}}\\
     \hline
     $q$ & Number of POW calls in a round for each party  \\ 
     $p$ & Probability of POW call to be successful and block mined  \\ 
     $\kappa$ & Length of hash, determines difficulty parameter $D = p 2^{\kappa}$  \\ 
    $n$ & Number of mining parties (we assume a flat setting) \\
    $t$ & Maximum number of parties controlled by the adversary \\
    $\delta$ & Honest Majority Parameter, $\delta \in (0,1)$ with $t \leq (1- \delta) (n-t)$ \\
    $\beta$ & Hashing power of the adversary per round, $\beta = t p q$ \\
    $\alpha$ & Hashing power of the honest parties per round, $\alpha = (n-t) p q$  \\
    $f$ & Total hashing power per round, $f = \alpha + \beta$ \\
    $\gamma$ & Probability that a round is successful $\gamma = 1 - (1-p)^{(n-t)q}$ \\
    $\gamma_u$ & Probability that a round is uniquely successful, $\gamma_u >(1-\frac{\gamma}{3}f)$ \\
    $\varepsilon$ & Typical execution parameter, $\varepsilon \in (0,1)$ \\
    $\lambda$ & Consecutive rounds needed for a typical execution \\  
    $c$ & Weight coefficient $c > 1$ and $c \in \mathbb{R}$ \\
    $K$ &  Weight parameter for the common weighted prefix property, $K \in \mathbb{R}$ \\
     \hline
    \end{tabular}
    \end{table}

\subsection{Typical execution}

We shall now introduce the formal notion of a \emph{typical execution}~\cite{DBLP:conf/eurocrypt/GarayKL15}, the idea is that if we have enough consecutive rounds, we can show that they fulfill certain properties with a high probability. Furthermore, we note that if we have have a set of consecutive rounds of a certain size, we can show that every subset of consecutive rounds within it, if large enough, also fulfills these properties. To define these properties we introduce the following notation, aligned with the Bitcoin Backbone paper~\cite{DBLP:conf/eurocrypt/GarayKL15}. 

We define $X_{ijk}$ to be a Boolean random variable that denotes whether in round $i$ the $j$-th query of the $k$-th honest party is successful.  Furthermore, let $Z_{ijk}$ be a Boolean random variable for the same case but for the $k$-th corrupted party mining. We also let $Y_i$ denote whether or not \emph{exactly one} honest party mines in round~$i$, and let $\tilde{X}_i$ represent whether or not any \emph{honest} party mines in round~$i$. A round with $Y_i = 1$ is called \emph{uniquely successful}. Given these, we define $X_i = \sum_{k=1}^{n-t} \sum_{j=1}^q X_{ijk}$ and $Z_i = \sum_{k=1}^{t} \sum_{j=1}^q Z_{ijk}$. For a set~$S$ of (consecutive) rounds, we define $X(S) = \sum_{r \in S} X_r$ and similarly for $Z(S)$, $\tilde{X}(S)$ and $Y(S)$. In summary, we obtain the following:

\begin{center}
    \small
  \begin{tabular}{|c|l|}
  \hline
   $X(S)$  & Total number of blocks mined by an honest party \\
  &  in consecutive rounds $S$.\\
   $\tilde{X}(S)$  & Total number of times an honest party  \\
  & mines in a round, for consecutive rounds $S$.\\
  $Z(S)$ &  Total number of blocks an adversary mines\\
  &  in consecutive rounds $S$. \\
  $Y(S)$ & Number of rounds in $S$ that are uniquely successful. \\
  \hline
  \end{tabular}
\end{center}

  We make the same \emph{honest majority assumption} as in the Bitcoin Backbone~\cite{DBLP:conf/eurocrypt/GarayKL15}, that there exists $\delta \in (0, 1)$ such that $t \leq (1- \delta) (n-t)$. Let also
  \begin{align*}
    \alpha &= \E[X_i] = p q (n-t)\\
    \beta  &= \E[Z_i]= t p q\\
    \gamma &= \E[\tilde{X}_i]
  \end{align*}
  from which it follows that
  $\E[Y]=\gamma_u=q(n-t)p(1-p)^{q(n-t)-1}>(1-\frac{\gamma}{3}f)$.  We
  assume that $3 \gamma + 3 \varepsilon < \delta \leq 1$, where $\gamma$ is
  the probability that a round is successful and $\varepsilon \in (0,1)$.

We also use Garay \emph{et al.}'s notions of insertions, predictions, and copies~\cite{DBLP:conf/eurocrypt/GarayKL15}. In particular, an \emph{insertion} occurs when, given a tree $T$ with two consecutive blocks $B$ and $B'$ a block $B^*$ created after $B'$ so that $B$, $B^*$, and $B'$ form three consecutive blocks of a valid chain inside the tree. A \emph{copy} occurs if the same block exists in two different positions in the tree. A \emph{prediction} occurs when a block extends one which was computed at a later round.

\begin{definition}\label{typical execution}
  An \emph{($\varepsilon, \lambda$)-typical execution} for $\varepsilon \in (0, 1)$ and $\lambda \geq 2/\gamma$, over a set $S$ of at least $\lambda$ consecutive rounds satisfies:
\begin{enumerate}
\item $(1- \varepsilon) \E[X(S)] < X(S) < (1 + \varepsilon) \E[X(S)]$
\item $(1- \varepsilon) \E[\tilde{X}(S)] < \tilde{X}(S) < (1 + \varepsilon) \E[\tilde{X}(S)]$
\item $(1 - \varepsilon) \E[Y(S)] < Y(S)$
\item $Z(S) < Y(S)$ and $ Z(S) < (1 + \varepsilon) \E[Z(S)]$ 
\item There are no insertions, predictions, or copies.
\item $\sum_{j=1}^q X_{ijk} \leq 1$ for every honest party $P_k$.
\end{enumerate}
\end{definition}

We note that the points (2)--(5) correspond to the conditions for a typical execution as defined by Garay \emph{et al.}~\cite{DBLP:conf/eurocrypt/GarayKL15}. 

\begin{restatable}{theorem}{negli}
  An execution is $(\varepsilon,\lambda)$-typical with probability \newline
   $ 1- e^{-\Omega(q \varepsilon^2 \gamma \lambda + \kappa + q)}$ .
\end{restatable}

\begin{proof}
    The proof is analogous to the proof in the Bitcoin Backbone paper~\cite{DBLP:conf/eurocrypt/GarayKL15}. It follows directly from applying a Chernoff bound to $X(S)$, $Z(S)$ and $\tilde{X}(S)$. We note $X_{ijk}$, $\tilde{X}_i$ and $Z_{ijk}$ are all independent Bernoulli trials. In all trials the probability that one of these is outside the given range is at most $2 e^{-\mu \varepsilon^2/3}$, where $\mu$ is the respective expected value. Garay \emph{et al.}~\cite{DBLP:conf/eurocrypt/GarayKL15} show that the expected values of these variables can all be rewritten to have an upper bound that is a factor of $\gamma$. Thus, an execution fulfills the first four criteria with probability $ 1- e^{-\Omega(q \varepsilon^2 \gamma \lambda)}$. They further showed that insertions, deletions and copies occur with probability  bounded by $ e^{-\Omega(\kappa)}$, as insertions and copies happen if a block extends two distinct blocks, which means a collision has occurred and a prediction occurs at an equally small likelihood.
    
    The final condition is directly influenced by the choice of $q$, and as $p$ is already small, a Chernoff bound can be used to show that this occurs with probability bounded by $ e^{-\Omega(q)}$.
    
    Using the Union bound, we combine the previous three bounds to finish the proof.
    \end{proof}

From now on, unless explicitly noted otherwise, all statements we make
assume the conditions of a typical execution hold.  In other words, we can
find parameters $\varepsilon$, $\gamma$, $\lambda$, $q$ and $\kappa$ so
that the properties hold with probability
$ 1- e^{-\Omega(q \varepsilon^2 \gamma \lambda + \kappa + q)}$.

\subsection{Properties of Medium}

For analyzing the protocol, we define some of its main properties in the
model of Garay \emph{et al.}~\cite{DBLP:conf/eurocrypt/GarayKL15}.

\begin{definition}[Normalized tree weight]
For a block $B$ in tree $T$, we define the \emph{normalized tree weight of $B$}, or $\bar{\omega}_c(T(B))$, to be the weight of the subtree on $B$ (or the tree weight of $B$) divided by the weight of $B$, or \[\bar{\omega}_c(T(B)) = \frac{\omega_c(T(B))}{c^{d(B)}}.\]
\end{definition}
\begin{definition}[$k$-dominant prefix]
We define the \emph{k-dominant} prefix of the chain $C$, or $C^{\lceil k}$, as the chain $C$ without any blocks $B$ for which $\omega_c(T(B)) < k$, with the parameter $k \in \mathbb{R}$. If there is no block $B$ in chain $C$ with $\tau_c(T(B)) \geq k$, $C^{\lceil k}$ is defined to be the genesis block.
  \end{definition}
  
We note that blocks are always removed from the head of the chain when computing the k-dominant prefix of a chain. We can now come to the properties.

\begin{definition}[Normalized tree weight growth]
  For parameters $\tau\in\mathbb{R},s\in\mathbb{N}$, for any honest block $B$ mined in round~$r$, and for a set of consecutive rounds $S$ with size $|S|=s$ starting just after round~$r$ it holds that when $B$ is in the main chain of every honest party $P_i$ during $S$, then the normalized weight of $B$ increases by at least weight $\tau$ in the local view of every honest party $P_i$.
\end{definition}

\begin{definition}[Chain growth]\label{def:chaingrowth}
  There exist parameters $g>0$ and $r_0\in\mathbb{N}$ such that in round $r\geq r_0$, every honest party adopts a chain of length at least~$g \cdot r$.
\end{definition}

\begin{definition}[Common weighted prefix]
  There exists a parameter $K \in \mathbb{R}$ so that for any pair of honest parties $P_1$ and $P_2$ that adopt main chains $C_1, C_2$ at rounds $r_1 \leq r_2$ in their respective local views, it holds $C_1^{\lceil K} \preceq C_2$.
\end{definition}

\begin{definition}[Fresh block]
  At round $r$ there exists a parameter $u \in \mathbb{N}$ so that for any subset $u$ consecutive rounds, there is at least one block mined by an honest party which is in the main chain of all honest parties in every round $r' \geq  r$.
\end{definition}

In the remainder of this section, we establish the chain growth, weight growth, common weighted prefix, and fresh block properties.  From these, it is possible to show that a robust public transaction ledger exists on top of our protocol, which satisfies liveness and persistence; we do this in the next section.

\subsection{Foundation lemmas and chain growth}

We use \emph{block trees} as defined by Kiayias and Panagiotakos~\cite{DBLP:conf/latincrypt/KiayiasP17}. $\CT_r^P$ is the tree formed from the blocks that honest party $P_i$ has received up to round $r$. $\CT_r$ is the tree containing \textit{all} blocks broadcast by any party up until round $r$. $\hat{\CT_r}$ is the tree that contains $\CT_r$ and also includes all blocks mined by honest parties at round $r$.
This means that for any honest party $P_j$, we have
\[ \CT_r^P \subseteq \CT_r \subseteq \hat{\CT_r} \subseteq \CT_{r +1}^P. \]
This follows intuitively from the fact that each honest party has a subtree of all broadcast blocks up to round $r$ in their local view at the start of round $r$, thus $\CT_r^P \subseteq \CT_r$. This subtree always contains all honest blocks broadcast in the previous round. As honest parties broadcast all newly mined blocks and blocks they received before round $r$ at the end of round $r$, $\CT_r \subseteq \CT_{r +1}^P$ must hold.

It is important to note the adversary can choose to only broadcast its blocks to certain honest parties, so two honest parties $P_1$ and $P_2$ may have received different blocks in round $r-1$, which means $\CT_r^{P_1} \ne \CT_r^{P_2}$. Thus the main chains of two honest parties may also differ in length.
 $\CT_r$ is the tree containing \textit{all} blocks broadcast by any party up until round $r$, the length of the main chain of this tree is unique, as there can only be multiple main chains in $\CT_r$ if each has the same length and weight. 

We define $\ell_{mc}(\CT_r)$ to be the length of the main chain in $\CT_r$. The length of the main chain in $\hat{\CT}_r$ is also unique (as honest parties extend the main chain by at most one block in a typical execution). As in $\CT_{r}^P$, the length of the main chain in $\CT_{r +1}^P$ is not necessarily unique.

The next remark introduces a different perspective that simplifies the upcoming proofs.
\begin{remark}\label{remark:chains}
  Given two chains $C_1, C_2$ in the local view of some honest party $P_i$, such that one of them is the main chain, w.l.o.g. $C_1$. The fact that $C_1$ is the main chain means that at some point in the chain $C_2$ there is a block $B_2$ that has a sibling $B_1\in C_1$ that has a heavier subtree. This follows directly from the fact that all the chains start with the genesis block and in every interaction the algorithm selects the block with the heaviest subtree.
\end{remark}

We shall start our analysis by discussing chain length growth behavior during a typical execution.

\begin{lemma}\label{HonestNodeMining}
If an honest party mines in round $r$ and the adversary does not broadcast in round $r-1$ it holds that
\[\ell_{mc}(\hat{\CT}_r) = \ell_{mc}(\CT_r) +1 .\]
Additionally, if this is an uniquely successful round all parties have the same local view and have the same main chain in $\hat{\CT}_r$.
\end{lemma}

\begin{proof}
This is clear from the protocol, honest parties always mine on the main chain, which is chosen by recursively selecting the block with the heaviest subtree and heaviest subtree resulting in the longest main chain if there are ties. Unless an adversary broadcasts in round $r-1$ all honest parties mine on the same main chain unless there was a block with more than one descendant that had a subtree of the same weight, resulting in two different main chains of the same length. Thus, if any honest parties are successful in round $r$, they extend the chain they are mining on by length 1 (only by length 1, due to point 5 of a typical execution (Definition~\ref{typical execution})). As there can only be multiple main chains in the local views of honest parties if they all have the same length any chain that is mined on in round $r$ by an honest party has the same length. Furthermore, the block that was mined in that round adds to the weight of the subtrees of all the previous blocks in the chain, thus a main chain in $\hat{\CT}$ is a chain that was mined on, which now has length $\ell_{mc}(\CT_r) +1$.

Furthermore, it is clear that if only one party mines, only one main chain is extended and thus there cannot be another main chain in the local view of an honest party as we have assumed the adversary has not broadcast in the round before. 
\end{proof}

With this we can prove the following lemma.

\begin{lemma}
  \label{NewLemma1}

Assume that an honest block $B_0$, mined in round $r_0$, stays in the main chain of all the honest parties for a set of consecutive rounds $S$ starting at round $r_0+1$, then the length increase of the main chain of a given party $P_i$, at the beginning of the first round just after $S$, is lower and upper bounded ($l(S)$ is the increase in length of the main chain during the set of rounds $S$) by:
\[Y(S)-Z(S)\leq l(S)\leq \tilde{X}(S)+Z(S).
  \]
In other words, the length increase is lower bounded by the number of uniquely successful rounds minus the number of adversarial blocks released in $S$, and, upper bounded by the number of successful rounds plus the number of adversarial blocks released in $S$.

\end{lemma}

\begin{proof}

  First of all, notice that only the blocks releases in the subtree of $B$ are relevant. Since all the honest parties agree that $B$ is in the main chain during all the execution, this means that blocks releases by the adversary mined previously to $B$ can safely be ignored.
  We analyze first the lower bound, $Y(S)-\hat{Z}(S)\leq l(S)$. The result follows by induction over the number of uniquely successful rounds $Y(S)$. First of all, notice that any adversarial block produced before round $r_0$ is completely irrelevant since the assumption is that $B_0$ remains in the main chain. In other words it is enough to analyze the structure of the subtree of $B_0$ and the adversarial blocks produced after or in rounds $r_0$.
  \begin{itemize}
    \item Case $Y(S)=0$, the bound is trivially satisfied.
    \item Case $Y(S)=1$. Since the hypothesis is that block $B_0$ stays in the main chain of any honest party, the unique uniquely successful block mined is a descendant of $B_0$. This implies that the main chain, which before the set of rounds $S$ finished in $B_0$, no longer finishes with $B_0$ ($B_0$ is no longer a leaf).
    \item Case $Y(S)=2$. This follows from the fact that $B_0$ stays in the main chain during all the execution and the existence of a chain of length two.
    \item Assume that the statement holds up to $n-1$. However, assume that the statement is not true for $n$. Precisely, denote by $r_1$ the round in which the last uniquely successful block of $S$ was mined, and define a set of rounds 
      \[
        S^{\prime}:=\{r\in\mathbb{N}|r_0< r < r_1 \}.
      \]
      We obtain a system of two equations,
      \begin{equation*}
        \begin{cases}
         l(S)<Y(S)-\hat{Z}(S)\\
         l(S^\prime) \geq Y(S^\prime)-\hat{Z}(S^\prime).
        \end{cases}
      \end{equation*}

      By definition of $S^\prime$, we observe that $Y(S^\prime)=Y(S)-1$ and $\hat{Z}(S^\prime)=\hat{Z}(S)-k$, where $k$ is the number of adversarial blocks released after the last uniquely successful block in $S$. Then we get
    \begin{equation*}
      \begin{cases}
       l(S)< Y(S)-\hat{Z}(S)\\
       l(S^\prime) \geq Y(S)-1-\hat{Z}(S)+k.
      \end{cases}
    \end{equation*} 

    Since the minimum increase in length is one, and negating the second inequality, this means
    \begin{equation*}
      \begin{cases}
       l(S)+1\leq Y(S)-\hat{Z}(S)\\
      - l(S^\prime) \leq -Y(S)+1+\hat{Z}(S)-k.
      \end{cases}
    \end{equation*}

    Adding both equations gives
    \[
      l(S)+1- l(S^\prime)\leq 1-k.
    \]
    and
    \[
      k\leq l(S)- l(S^\prime).
    \]
    This is a contradiction since $k\geq 0$.  Thus, $ l(S)\geq l(S^\prime)$
    and we see that
    \begin{equation*}
      \begin{cases}
       l(S)< Y(S)-\hat{Z}(S)\\
       l(S)\geq l(S^\prime) \geq Y(S)-1-\hat{Z}(S)+k.
      \end{cases}
    \end{equation*}
    Taking into consideration that the minimum increase in length is one and that $k$ is non-negative, it holds
    \begin{equation*}
      \begin{cases}
       l(S)< Y(S)-\hat{Z}(S)\\
       l(S)>Y(S)-\hat{Z}(S)
      \end{cases}
    \end{equation*}
    We conclude that the statement holds for $Y(S)=n$.
    This completes the inductive step and proves the lower bound in the lemma.
  \end{itemize}

  The upper bound follows trivially from the fact that the best case for length growth is when the adversary collaborates with the honest parties. 
\end{proof}

\begin{lemma}
  Assume a set of consecutive rounds $S$ after an honest $B_0$ is mined in round $r_0$, with $|S| \geq \lambda$. Assume $B_0$ is part of the main chain during the set of rounds $S$, then the length increase of the main chain f any honest party $P_i$ is lower bounded by 
\[(1-\varepsilon)\E[Y(S)]-(1+\varepsilon)\E[Z(S).]\]
\end{lemma}
\begin{proof}
  Follows from Lemma~\ref{NewLemma1} and the properties of an ($\varepsilon, \lambda$) -typical execution.
\end{proof}

\begin{corollary}[\rm\em Chain growth]\label{ChainGrowth}

The chain growth property (Definition~\ref{def:chaingrowth}) holds with parameters $r_0=\lambda$ and $g= (1-\varepsilon)\gamma_u-(1+\varepsilon)\beta$.
\end{corollary}

\begin{proof}
We apply Lemma \ref{NewLemma1} together with the fact that the genesis block is always part of the main chain for every honest party and $S=\{r\in\mathbb{N}|r^\prime\leq r\}$ satisfies that $|S| \geq \lambda$. Thus, the conditions of an ($\varepsilon, \lambda$)-typical execution hold. These conditions also imply that $((1-\varepsilon)\gamma_u-(1+\varepsilon)\beta)\cdot r=(1-\varepsilon)\E[Y(S)]-(1+\varepsilon)\E[Z(S)]>0$, where $S$ is the set of rounds until $r$.
 \end{proof}

\subsection{Weight growth property}\label{Section:growth}
In this section we commence the full analysis to prove that Medium satisfies the normalized tree weight growth property. 

We introduce notation needed to formalize bounds on the weight increase of the blocks in the main chain.

\begin{definition} Given a block $B$ and a round $r$ such that $B$ is in the main chain of some honest party $P_i$ we define \[\ell_B := \min_{P_i \text{honest}}\{ \ell_{mc}(\CT_{r}^P) \} - d(B).\]
$\ell_B$ is the minimum distance from block $B$ to the head of the main chain in the local view of party $P_i$. 
We define $\ell_L$ as the maximal distance from a block $B$ in the main chain of any honest party to any head of a chain in its subtree.
\end{definition}

\begin{lemma}\label{TreeWeightBounds}

For any honest block $B$ mined in round $r$ and any set of consecutive rounds $S$ starting just after $r$, consisting of at least $\lambda$ rounds, the normalized weight increase $\Delta \bar{\omega}_c(T(B))$ respects
\[
  \Delta \bar{\omega}_c(T(B))  < \sum_{i=1}^{\lceil (1+\varepsilon)(\gamma+\beta)|S| \rceil} k(i,|S|)\ c^{i}.
\]
Where
\[
  k(i,|S|)= \left\{ \begin{array}{lcc}
    1 &   \text{if}  & i<\lceil (1+\varepsilon)(\gamma+\beta)|S| \rceil- \big\lceil\frac{(1+\varepsilon)(\alpha-\gamma)|S|}{(1+\varepsilon)\gamma}\big\rceil\\
    \\ (1+\varepsilon)\alpha &  \text{if} &i\geq\lceil (1+\varepsilon)(\gamma+\beta)|S| \rceil- \big\lceil\frac{(1+\varepsilon)(\alpha-\gamma)|S|}{(1+\varepsilon)\gamma}\big\rceil
    \end{array}
\right.
.
\]

If $B$ is in the main chain of every honest party during $S$, then the normalized weight increase is also lower bounded by
\[
    \sum_{i=1}^{\lfloor ((1-\varepsilon)\gamma_u-(1+\varepsilon)\beta)|S| \rfloor} c^i<\Delta  \bar{\omega}_c(T(B)).  
\]
Both bounds in the local view of any honest party $P_i$.
\end{lemma}

\begin{proof}
On the one hand, the maximum weight increase occurs when the adversary collaborates with the honest parties and the honest parties mine in the subtree of $B$, and both the adversary and the honest parties succeed as often as possible. This respects the conditions of an $(\varepsilon,\lambda)$-typical execution and we can apply Lemma~\ref{NewLemma1}.

From Lemma~\ref{NewLemma1}, there is an upper bound in the length increase of the main chain of any honest party that considers $B$ as part of the main chain: $l(S)\leq\tilde{X}(S)+\hat{Z}(S)<(1+\varepsilon)(\gamma+\beta)|S|$, using the conditions of an $(\varepsilon,\lambda)$-typical execution. Furthermore, the weight is maximized when all the forked blocks occur as deep as possible in the tree. Again, by the properties of an $(\varepsilon,\lambda)$-typical execution, the number of blocks in the tree is bounded by $(1+\varepsilon)(\alpha+\beta)|S|$, and the number of honest blocks mined per round is upper bounded by $(1+\varepsilon)\alpha$. We have at most $(1+\varepsilon)(\alpha-\gamma)|S|$ forked blocks, and in every level of the tree up to $(1+\varepsilon)\gamma$ blocks. We conclude that the best case for weight increase occurs when the last $\lceil\frac{(1+\varepsilon)(\alpha-\gamma)|S|}{1+\varepsilon)\gamma}\rceil$ levels of the tree contain all the forked blocks. Defining $k(i,|S|)$ as in the statement,
\[
  k(i,|S|)= \left\{ \begin{array}{lcc}
    1 &   \text{if}  & i<\lceil (1+\varepsilon)(\gamma+\beta)|S| \rceil- \big\lceil\frac{(1+\varepsilon)(\alpha-\gamma)|S|}{1+\varepsilon)\gamma}\big\rceil\\
    \\ (1+\varepsilon)\alpha &  \text{if} &i\geq\lceil (1+\varepsilon)(\gamma+\beta)|S| \rceil- \big\lceil\frac{(1+\varepsilon)(\alpha-\gamma)|S|}{1+\varepsilon)\gamma}\big\rceil
    \end{array}
\right.
.
\]
Hence,
\[\Delta \bar{\omega}_c(T(B)) < \sum_{i=1}^{\lceil (1+\varepsilon)(\gamma+\beta)|S| \rceil} k(i,|S|)\ c^{i}.
  \]

On the other hand, the minimum weight increase occurs when the main chain of some party $P_i$ whose local view includes $B$ in the main chain, is as short as possible. From Lemma~\ref{NewLemma1}, we observe that $l(S)\geq Y(S)-Z(S)$, and using the conditions of an $(\varepsilon,\lambda)$-typical execution, it follows $l(S)>((1-\varepsilon)\gamma_u-(1+\varepsilon)\beta)|S|$. Furthermore, the worst case for the weight increase is when the adversary achieves this with allowing any superfluous block to the subtree of $B$. Hence

\[
  \Delta \bar{\omega}_c(T(B)) > \sum_{i=1}^{\lfloor ((1-\varepsilon)\gamma_u-(1+\varepsilon)\beta)|S| \rfloor} c^i.
\]
\end{proof} 

We can now apply these bounds to achieve the normalized tree weight growth property.

\begin{theorem}[\rm\em Normalized tree weight growth]

The normalized tree weight growth property holds with parameters $s = |S| \geq \lambda$ and 

\[
  \tau=\sum_{i=1}^{\lfloor ((1-\varepsilon)\gamma_u-(1+\varepsilon)\beta)|S| \rfloor} c^i. 
\]
\end{theorem}

\begin{proof}
This follows directly from Lemma \ref{TreeWeightBounds}.
\end{proof}

\subsection{Common weighted prefix property}

Lemma~\ref{TreeWeightBounds} has further applications than the ones discussed in Section~\ref{Section:growth}.
\begin{remark}\label{WeightIncrease}
Any block $B$ requires at least $\lambda$ consecutive rounds to get a normalized subtree-weight of at least 
\[ \sum_{i=1}^{\lceil (1+\varepsilon)(\gamma+\beta)\lambda \rceil} k(i,\lambda)\ c^{i}.\]
This is a direct consequence of Lemma~\ref{TreeWeightBounds}.
\end{remark}

This inequality constitutes the baseline to our proof of the common weighted prefix property. Before we go into this, we introduce two complementary lemmas.

\begin{restatable}{lemma}{case}\label{Case3}

Assume there exists a fork in $\CT_r^P$, where $P_i$ is any honest party. Denote by $C_1$ and $C_2$ two unique chains produced by this fork, which have the same prefix prior to this fork; assume that the last block in this common prefix was mined in round $r' \leq r$. Denote by $B_1$ and $B_2$ the first block in each chain after the fork. Further assume $\omega_c(\CT_r^P(B_1)) < \omega_c(\CT_r^P(B_2))$, $\ell(C_1) = \ell(C_2) + s$ for $s > 0$. Then, the adversary had to release $s$ blocks from round $r'-1$ to $r-1$.
\end{restatable}

 \begin{proof}
    To prove this statement we assume that there are no blocks in the subtree $\CT_r^P(B_2)$ that are at a greater depth than the length of $C_2$. Still assuming this assumption holds let now assume the statement of the lemma does not hold and find a contradiction.
    
    Take any block $B_i$ in $C_1$ at depth $\ell(C_2) +i$, for $i \in [1, s]$, this block was mined at $ r' \leq r_i \leq r$. We shall show that for each of the blocks $B_i$ the adversary had to release at least one block for there to exist a tree of such a shape at round $r$.
    
    If $B_i$ is corrupted we do not have to show anything, thus we assume $B_i$ is honest. This means for at least one honest party $P'$ \[\omega_c(\CT_{r_i}^{P'}(B_1)) \geq \omega_c(\CT_{r_i}^{P'}(B_2)). \] 
     Now, since we know that in round $r$
    \[  \omega_c(\CT_r^P(B_2)) > \omega_c(\CT_r^P(B_1)) > \omega_c(\CT_r^P(B_1)) \geq \omega_c(\CT_{r_i}^{P'}(B_1)) + \sum_{j=i}^{s} c^{\ell(C_2) +j}\]
    we know that between rounds $r_i-1$ and $r$ there must have been at least $s-i+1$ blocks mined on $C_2$. (This follows from $\sum_{j=0}^{s-i} c^{s-j} > s- i+1$). 
    \newline
    
     We examine two cases.
    \begin{enumerate}
    \item  The first is, if the adversary did not broadcast in round $r_i -1$. This means all parties have the same local view of  $\CT_{r_i}$. Thus, an honest party can only mine on $C_2$ (to produce the missing $s-i+1$ blocks on $C_2$) at round $r_i$, if $\CT_{r_i}(B_1)$ and $\CT_{r_i}(B_2)$ have the same weight and result in main chains of the same length, but this contradicts $B_i$ being the only block at this depth . Thus, the blocks must have been mined after round $r_i$, but after round $r_i$, $B_1$ is the sibling with the heaviest tree weight, thus no honest party would have mined on $C_2$. Thus, the adversary must have released $s-i+1$ blocks on the subtree of $B_2$ for this fork to occur, or, if $i \ne s$ it can switch the local view of another honest party before further blocks are released on $C_1$. If it does this after round $r_i$ it needs to compensate the weight produced on $C_1$ in this round and has to release more than one block (as blocks in the subtree on $B_2$ have a strictly lower depth). Otherwise the adversary could have changed the local view of another honest party before $r_i$,  this is the second case.
    \item If the adversary did broadcast in round $r_i -1$ it is possible to create two different local views by only releasing blocks to certain honest parties and have honest parties mine blocks on $C_2$ in that round. As the adversary must expend at least a block for this it just remains for us to show it cannot 'compensate' multiple $B_i$ in this manner. (We note that under the conditions of an $(\varepsilon,\lambda)$-typical execution honest parties do not mine on their own blocks during a round and can only extend the length of a chain by 1 block). 
    
    If honest parties release enough blocks for $B_2$ to have a heavier tree weight than $B_1$ after round $r_i$, i.e. $\omega_c(\hat{\CT}_{r_i}(B_1)) < \omega_c(\hat{\CT}_{r_i}(B_2))$ we note honest parties do not mine on $C_1$ without the adversary releasing further blocks. If we would like the following $B_{i+1}$ to be honest we come back to this case, if $i=s$ we are done. Alternatively, the honest parties could mine enough weight for $B_1$ to have equal or heavier tree weight than $B_2$ after round $r_i$, if there are less than $k < s-i$ corrupted blocks needed for $B_2$ to have a higher tree weight than $B_1$ in round $r$ than the adversary has won. As we assumed blocks on the subtree of $B_2$ cannot weigh more than $c^{\ell(C_1)}$ this leads us to the condition that $k > \sum_{j = i+1}^s c^j$ (as further blocks must be released on $C_1$), and thus $k > s-i-1$, which would be a contradiction. 
    \end{enumerate}
    
    It remains to show this still holds if there are blocks in the subtree $\CT_r^P(B_2)$ that are at a greater depth than the length of $C_2$. We show this follows recursively from our statement. If there was a block in $\CT_r^P(B_2)$ at a greater depth there would be a fork in this subtree with blocks $B_3$ and $B_2^*$  broadcast in round $r^* \geq r'$ resulting in two chains, $C_2$ and another $C_3$ with $\ell(C_3) = \ell(C_2) + s^*$, $s^* > 0$ and $\CT_r^P(B_2^*) > \CT_r^P(B_3)$. We apply this until there are no blocks in the heavier subtree with a longer length than the length of the main chain and then apply our proof. Then, such a subtree is only possible if the adversary broadcast at least $s^*$ blocks from round $r^* -1$ to $r-1$. Using this our proof still holds.
    
    Thus, for every block $B_i$ in $C_1$ at depth $\ell(C_1) +i$, for $i \in [1, s]$ there is a corresponding corrupted block and the adversary must broadcast at least $s$ blocks to produce such a fork.

    \end{proof}
  
We are now able to discuss the behavior of chains when removing blocks of a specific tree weight, we shall show the weighted common prefix property must hold by proving the following lemma. As we must take multiple cases of different possible tree structures into account the proof is quite lengthy.

\begin{restatable}{lemma}{CommonPrefixLemma}\label{Weighted Common Prefix Lemma}
  Suppose at round $r$ of an ($\varepsilon, \lambda$)-typical execution, an honest party has a chain $C_1$ and a chain $C_2$ is adopted by an honest party, such that $C_2$ differs from $C_1$ in a block $B_2$ with $\omega(T(B_2))\geq \omega(T(B_1))$. That is, the blocks before $B_2$ in $C_2$ are the same as in $C_1$ and $C_1$ has $B_1$ in the place of $B_2$.  Then $C_1^{\lceil K} \preceq C_2$ and  $C_2^{\lceil K} \preceq C_1$ for weight
  \[
    K = \sum_{i=1}^{\lceil (1+\varepsilon)(\gamma+\beta)\lambda \rceil} k(i,\lambda)\ c^{i}.
  \]
\end{restatable}

\begin{proof}

    We assume by contradiction, either $C_1^{\lceil K} \npreceq C_2$ or $C_2^{\lceil K} \npreceq C_1$. Consider the last block of the common prefix of $C_1$ and $C_2$ that was computed by an honest party at round $r^*$ and at depth $\ell$ (this block could be genesis). We define $S = \{ i : r^* \leq i \leq r\}$ and note that $|S|\geq\lambda$, due to Lemma~\ref{TreeWeightBounds}. We shall show that this implies $Z(S) \geq Y(S)$ which is a contradiction. (We note that $Z(S) \geq Y(S)$ is only dependent on the size of $S$, thus $Z(S') \geq Y(S)$ holds for $S'$ if $|S'| = |S|$, here we define $S' = \{ i : r^* -1\leq i \leq r-1\}$. )
    
    To do this we shall examine an injection between the uniquely successful rounds in $S$ (their number given by $Y(S)$) and the blocks needed to ``balance them'' on the other chain, so that at round $r$ two different honest parties can have two different local views of the main chain. 
    
    We look at a uniquely successful round $r_i$, where the honest party who mined $B_i$ at this round mines on the main chain in their current local view. We look at 3 different cases for the main chain $C$ at $\CT_{r_i}^P$ that the honest party $P_i$ mines on. We first assume the honest party mines on chain $C_1$ or $C_2$, without loss of generality we assume that the honest party is mining on chain $C_1$, by $\ell_i(C_1)$ we denote the length of the chain at that round in the local view of the honest party mining in that round.
    
    \begin{itemize}
    \item \textbf{Case 1:} $\ell_i(C_1) > \ell_i(C_2) $ to balance this block on $C_2$ the adversary must release more than one block on the other side, due to blocks at lower depths having more weight.
    \item \textbf{Case 2:} $\ell_i(C_1) = \ell_i(C_2) $ to balance this block on $C_2$ the adversary must release one block at that level or more (if there is already a weight difference, or it cannot mine a block at the depth)
    \item \textbf{Case 3:} $\ell_i(C_1) < \ell_i(C_2) $ from Lemma~\ref{Case3} we know that for $\CT_{r_i}^P(B_1)$ to weigh more than $\CT_{r_i}^P(B_2)$ (and have a common root produced in round $r^*$) but be shorter by a length of $s>0$  the adversary must have already broadcast $s$ parties, even if in the best case the
    \[ c^{s+1} > \sum_{i=1}^s c^i = \frac{(c^s -1) c}{(c-1)}\] which holds for $c<2$ and the adversary can 'balance' $s$ blocks with 1 block, he must still broadcast $s$ blocks to produce this kind of subtree, and thus still needs at least as many corrupted blocks as uniquely successful rounds to create this fork. 
    \end{itemize}
    In all these cases we see that $Z(S) \geq Y(S)$ must occur for the adversary to win, which contradicts the assumption of a typical execution. 
    
    We still need to review what happens when the the honest party $P_i$ mines on a different chain than $C_1$ or $C_2$, we call this chain $C_3$, for this to happen $C_3$ must be the main chain in its local view at the uniquely successful round $r_i$ and there are blocks $B_{3,1}$ and $B_{3,2}$ so that $\omega_c(\CT_{r_i}^P(B_{3,1})) \geq \omega_c(\CT_{r_i}^P(B_1'))$  and $\omega_c(\CT_{r_i}^P(B_{3,2})) \geq \omega_c(\CT_{r_i}^P(B_2'))$ where $B_{3,1}$ is the first block on $C_3$ after it forks from $C_1$ and $B_{3,2}$ the first after $C_3$ forks from $C_2$, the blocks $B_1'$ and $B_2'$ are the first blocks in $C_1$ and $C_2$ after the fork of their respective chains from $C_3$. 
    \begin{itemize}
    \item  \textbf{Case 1:} $\ell_i(C_3) \geq \ell_i(C_2) $ and $\ell_i(C_3) \geq \ell_i(C_1) $ to balance this block on $C_2$ and $C_1$ the adversary must release at least one block on \emph{both} $C_2$ and $C_1$.
    \item  \textbf{Case 2:} $\ell_i(C_3) \geq \ell_i(C_2) $ and $\ell_i(C_3) <  \ell_i(C_1) $, or $\ell_i(C_3) \geq \ell_i(C_1) $ and $\ell_i(C_3) <  \ell_i(C_2) $. For $\ell_i(C_3) \geq \ell_i(C_2) $ resp. $\ell_i(C_3) \geq \ell_i(C_1) $ at least one block has to be released on $C_2$ resp. $C_1$ to balance the block on $C_3$. How many blocks were needed to produce $\ell_i(C_3) <  \ell_i(C_1) $ while $\omega_c(\CT_{r_i}^P(B_{3,1})) > \omega_c(\CT_{r_i}^P(B_1'))$ is slightly more complicated. To apply Lemma \ref{Case3} we must go back to the round where the root of $C_3$ and $C_1$ entered the block tree, which could be before $r^*$. Therefore, we observe the following, $C_1$ is at least $s_1 >0$ longer than $C_3$, as we are in an $(\varepsilon,\lambda)$-typical execution insertions do not occur, therefore the rounds these blocks were mined in must have been after round $r^*$. We know the adversary must have broadcast at least $s_1$ blocks from rounds $r^* -1$ to $r_i-1$ for such a tree structure to exist, the same is true for the case where $\ell_i(C_3) <  \ell_i(C_2) $.
    \item  \textbf{Case 3:} $\ell_i(C_3) < \ell_i(C_2) $  and $\ell_i(C_3) <  \ell_i(C_1) $ as already shown this means that for both sides the length difference $s > 0$ must be produced earliest at round $r^*-1$ by at least an equal number of blocks.
    \end{itemize}
    In all these cases we deduce that $Z(S') \geq Y(S)$ this contradicts the assumption of an $(\varepsilon,\lambda)$-typical execution and we have proved the weighted common prefix lemma.
  \end{proof}

From this, the common weighted prefix property follows directly. 
\begin{theorem}[\rm\em Common weighted prefix]\label{Common Weighted Prefix Property}
  Let
  \begin{equation}
    K = \sum_{i=1}^{\lceil (1+\varepsilon)(\gamma+\beta)\lambda \rceil} k(i,\lambda)\ c^{i}
    \label{Kdefinition}
  \end{equation}
  be the normalized tree weight. Then, for any pair of honest parties $P_1$
  and $P_2$ adopting chains $C_1$ and $C_2$ at rounds $r_1 \leq r_2$ in
  their respective local views, respectively, it holds
  $C_1^{\lceil K} \preceq C_2$.
\end{theorem} 

\begin{proof}

We assume the theorem is not true and find a contradiction. This means that there are rounds $r_1 \leq r_2$ where honest parties $P_1$ and $P_2$ adopt chains $C_1$ and $C_2$ as their main chains respectively but $C_1^{\lceil K} \npreceq C_2$. From Lemma \ref{Weighted Common Prefix Lemma} we know that for all chains $\tilde{C}_i$ in the local view of an honest party in round $r_1$ it must hold that $C_1^{\lceil K} \preceq \tilde{C}_i$ and $\tilde{C}_i^{\lceil K} \preceq C_1$ . It follows that $\tilde{C}_i \npreceq C_2$, which means that $C_2$ is not an extension of a chain that was in the local view of an honest party at round $r_1$.

This means that there must exist a round $r \geq r_1$ where an honest party adopted a chain $C'$ over a chain $C$ s.t. $C_1^{\lceil K} \preceq C$ but $C_1^{\lceil K} \npreceq C'$ (implied by $C_1^{\lceil K} \npreceq C_2$) , however in this round we can again apply Lemma \ref{Weighted Common Prefix Lemma} that $C^{\lceil K} \preceq C'$, furthermore, as $C$ is an extension of a chain that was in the main chain of an honest party at round $r_1$ we know that since blocks in the chain could not decrease weight, thus $C_1^{\lceil K} \preceq C^{\lceil K}$ must hold, which implies $C_1^{\lceil K} \preceq C'$ and is a contradiction. Thus we have proved the common weighted prefix property.
\end{proof}

It still remains for us to show the fresh block property, which we shall prove with our own version of the Chain Quality Lemma.

\subsection{Fresh block property}

Finally we prove that honest blocks eventually enter the ledger.

\begin{theorem}[\rm\em Fresh block]
  \label{Fresh block}
The fresh block property is satisfied with parameter \begin{equation} u = \frac{\hat{R}^2 + 2 \hat{R}}{(1 - \varepsilon) 2 \gamma} + \lambda R\end{equation}\label{uDefinition}
where $\hat{R}= \lfloor \frac{R}{2}\rfloor$ and $R$ is the maximal constant that fulfills the equation
\[ \sum_{i=1}^{R+1} c^i \leq  \sum_{i=1}^{\lceil (1+\varepsilon)(\gamma+\beta)\lambda \rceil} k(i,\lambda)\ c^{i}.\]
\end{theorem}

\begin{proof}
We use a similar proof strategy as the one applied for the common weighted prefix property~\ref{Common Weighted Prefix Property}. We analyze the honest blocks produced in a successful round during these $u$ consecutive rounds, and show that an honest block mined in a successful round enters the main chain and remains there in all subsequent rounds. This is proven thanks to the constrains of the number of corrupted blocks produced by the adversary in an $(\varepsilon,\lambda)$-typical execution.

However, the structure of the tree at the start of these $u$ rounds plays a role in how many honest blocks can be 'balanced' by the adversary. Assume the best case for the adversary that there exists another chain that is $R$ blocks longer than the current main chain, this chain differs from the main chain at a fork produced in round $r$ (this could be prior to the start of the $u$ rounds), so that the weight of the subtree at $B_{mc}$ (that results in the main chain) must be greater than the weight of the subtree on $B_R$ (that results in the $R$-blocks-longer chain). By releasing a block at depth $R + 1$, the adversary can compensate up to $H$ honest blocks, where $H$ is the largest constant that satisfies
\[  \sum_{i=1}^H c^i \leq c^{R+1}. \]
We note $H$ is at maximum $R$. Whenever such a chain exists it is a vulnerability. However, we can show the length of such a chain is bounded in an $(\varepsilon,\lambda)$-typical execution, if
\[ \sum_{i=0}^{R} c^i > K = \sum_{i=1}^{\lceil (1+\varepsilon)(\gamma+\beta)\lambda \rceil} k(i,\lambda)\ c^{i}\]
the common weighted prefix property no longer holds, as if the adversary released a block at depth $R+ 1$ only to certain honest parties then they would adopt this longer chain as their main chain. However, the $K$-prefix of this chain is not an extension of the other main chain, which is a contradiction. Thus, the maximal possible length of such a chain is bounded by the maximal constant $R$ that solves the equation$ \sum_{i=0}^{R} c^i \leq  K$.
Furthermore in an $(\varepsilon,\lambda)$-typical execution at the start of these $u$ consecutive rounds there can be no chain that is longer than $R$. 

After compensating $H$ honest blocks, mined in uniquely successful rounds by one corrupted block, in an $(\varepsilon,\lambda)$-typical execution the adversary has at most $H-2$ blocks left, which it can use to build a fork with a chain that is $H-2$ longer than the current main chain and compensates a certain number of honest blocks $H'$ (at maximum $R-2$), mined in uniquely successful rounds. Using the additional corrupted blocks mined during these rounds the adversary can build a fork with a $H'-2$ longer chain, this can continue until the adversary has used all the 'additional' blocks. In the worst case it takes 
\[ \sum_{i = 0}^{\hat{R}} (2i+1) = \hat{R}^2 + 2 \hat{R}\] uniquely successful rounds, with $\hat{R}= \lfloor \frac{R}{2}\rfloor$,  for all the adversary's additional blocks to be used. After this point the adversary always has to release strictly more than one block to compensate the weight produced in each uniquely successful round to prevent the block mined in that round from entering the main chain. From the properties of an $(\varepsilon,\lambda)$-typical execution, we know that in $u - R \lambda$ rounds, there are at least $\hat{R}^2 + 2 \hat{R}$ uniquely successful rounds, as 
\[
  (1 - \varepsilon) (u -R \lambda) \gamma = \hat{R}^2 + 2 \hat{R}.
\]
Moreover, there is at least one honest block entering the main chain. 

Enough blocks must now be mined on this honest block for it to have a large enough tree weight for it to be in the $K$-prefix of the main chain. As the adversary has no 'additional' blocks in every successful round it must release at least one block to prevent that block from entering the main chain. We consider groups of $\lambda$ blocks.
\begin{itemize}
  \item In the first group one honest block (more specifically $A :=\alpha \lambda (1 - \epsilon) - \beta \lambda (1 + \epsilon)$ blocks) enters the main chain.
  \item In the second group, if the adversary removes the extra block(s) from the previous group, $2A$ honest blocks  enter.
  \item In the $(R+1)$-th group, $(R+1) A$ blocks  enter the main chain, thus the first of these blocks has a normalized tree weight of at least  $\sum_{i=0}^{(R+1)A -1} c^i$, which is greater than $K$ and thus enters in the $K$-prefix and is stable.
\end{itemize} 

\end{proof}

\section{Robust public transaction ledger}
After discussing the security analysis of the protocol and the properties that Medium satisfies, we prove that Medium also constitutes a solid basis for a robust public transaction ledger.

\begin{definition}
  \label{def:stable}
  We define a transaction to be \emph{stable} if it is included in a block that is in the $K$-prefix of the main chain $C$ of an honest party at a given round $r$, where $K$ is the parameter defined in Equation (\ref{Kdefinition}).
\end{definition}
By the properties of the $K$-prefix of the main chain (e.g. Theorem~\ref{Common Weighted Prefix Property}), this implies that a stable transaction in round $r$ is also stable for any round $r^\prime>r$ and for any honest party.

It now only remains to show that a public transaction ledger which implements the Medium protocol satisfies liveness and persistence. We use the definitions given by Kiayias and Panagiotakos~\cite{DBLP:conf/latincrypt/KiayiasP17} which have been shown to satisfy the conditions needed for a robust transaction ledger. 

\begin{itemize}
\item \textbf{Persistence} holds if in round $r$ an honest party reports a transaction as stable, then whenever another party reports it as stable it remains in the same position in the transaction ledger.
\item \textbf{Liveness} holds if when a transaction is given as an input to all honest party for $u \in \mathbb{N}$ rounds then all honest parties eventually report this transaction as stable.
\end{itemize}

\begin{lemma}[\rm\em Persistence]
If a transaction is included in a block at position $k$ in the stable portion of the main chain of an honest party, i.e. in $C^{\lceil K}$, then when it enters the stable portion of the main chain of another honest party, it is located at the same position.
\end{lemma}

\begin{proof}
This follows directly from the common weighted prefix property (Theorem~\ref{Common Weighted Prefix Property}). If the transaction were at a different position for any other honest party this would mean that the two chains would not have the same prefix and this would be a contradiction to the common weighted prefix property. 
\end{proof}

\begin{lemma}[\rm\em Liveness]
  If a transaction is given repeatedly as input to all honest parties for $u$ consecutive rounds, then all honest parties  eventually report the transaction as stable.
\end{lemma}

\begin{proof}

This follows directly from the fresh block property (Theorem~\ref{Fresh block}).  It states that for every $u$ rounds an honest block is taken up into the main chain and stays in the main chain for all the following rounds, for all the honest parties. In other terms, this property guarantees that if a transactions is given to honest parties for $u$ rounds, it is taken up into the main chain and reported as stable by all the honest parties by the end of those $u$ rounds.

\end{proof}

\section{Throughput}

The particular characteristics of Bitcoin, GHOST, and Medium allow to compare the throughput of these protocols in a unified and simplified manner. Namely, all protocols select one main chain as the correct one and ignore every block that is not part of it. 

Bagaria \emph{et al.}~\cite{DBLP:conf/ccs/BagariaKTFV19} show that for Bitcoin, throughput is bounded by a security constraint which ensures that the malicious chain cannot grow faster in expectation than the honest main chain,
\begin{equation}
  \label{throu_bitcoin}
 \alpha (1 - \psi_f) > \beta.
\end{equation}
The variable $\psi_f$ stands for the probability that a block forks, i.e. the probability that a successful round is not uniquely successful. Without this constraint,, an adversary would be able to build a secret chain that eventually becomes longer than the main chain of any honest party.
It is clear that the throughput of the Bitcoin protocol is limited when this probability is small. A low forking probability is correlated with a low mining ratio (number of blocks mined per unit of time). Therefore, Bitcoin's throughput is limited by this constraint. 

In the case of Medium, since the weight of a block increases exponentially with its depth in the tree, one might suspect that the security constraint is the same as in Bitcoin. However, this is not exactly the case.

On the one hand, if a hypothetical adversary had access, for unlimited time, to some set of corrupted parties, the above constraint~(\ref{throu_bitcoin}) still applies. The reason for this is that despite the contribution of the forked blocks, the secret chain of the adversary becomes at some point long enough to compensate for this.

On the other hand, if we consider a more realistic scenario, in which the adversary is allowed to perform this attack for some set of consecutive rounds rounds $S$ only, the result is slightly different.

\begin{lemma}\label{expectedTreeWeight}
  The expected weight of a subtree with $N$ blocks and depth $\ell$, starting at depth $\ell_0$ produced by honest parties running the Medium protocol is
  $  \frac{N}{\ell} \sum_{i=1}^{\ell} c^{i + \ell_0}$.
\end{lemma}
\begin{proof}
  When honest parties run the protocol, the main chain is always the longest and they only split mining power when a fork occurs. In a given round, the probability of multiple honest parties mining is constant. If all parties mine on a chain at depth~$d$, the probability that there are multiple blocks mined at depth $d$ is given by this constant. When only honest parties mine, after a successful round parties always increase the depth they are mining at, thus at every depth the probability of there being multiple blocks is constant. 
    
    A subtree of depth $\ell$ has at least weight $\sum_{i=1}^{\ell} c^{i + \ell_0}$, the tree is rooted at depth~$\ell_0$. The rest of the blocks, in total $N-\ell$, can be at any depth in the subtree, therefore they follow a uniform distribution. This means that the expected weight is $\frac{N-\ell}{\ell} \sum_{i=1}^{\ell} c^{i + \ell_0}$ for these blocks, adding this to the weight of the previous blocks gives us the expected weight of the subtree.
    \end{proof}

    From now on, assume that the adversary builds a secret chain after some honest block $B_0$, and all the weights are normalized by $c^{\var{depth}(B_0)}$. Writing $s=|S|$, the expected value of honest (malicious) blocks in a set of consecutive rounds $S$ is $\alpha s$ ($\beta s$, respectively).
    The relative weight of this secret chain ($C_s$) is
\[
    \sum_{i=1}^{\lceil\beta s\rceil} c^i=\frac{c(c^{\lceil\beta s\rceil}-1)}{c-1}\simeq \frac{c(c^{\beta s}-1)}{c-1}. 
\]

Regarding the honest subtree, its expected number of blocks is $\alpha s$ and its depth $\alpha s (1-\psi_f)$, since a block does not fork, and increases the depth of the subtree, with probability $1-\psi_f)$. Using Lemma~\ref{expectedTreeWeight}, the expected weight of this subtree can be written as

\begin{align*} 
  \frac{\lfloor\alpha s\rfloor}{\lfloor\alpha s (1-\psi_f)\rfloor}\sum_{i=1}^{\lfloor\alpha s (1-\psi_f)\rfloor} c^i
  & \geq \frac{\alpha s}{\alpha s (1-\psi_f)}\sum_{i=1}^{\lfloor\alpha s (1-\psi_f)\rfloor} c^i\\
  & \geq \frac{1}{(1-\psi_f)}\sum_{i=1}^{\lfloor\alpha s (1-\psi_f)\rfloor} c^i
\end{align*}
\begin{align*}
   \mbox{} &> \sum_{i=1}^{\lfloor\lfloor\alpha s (1-\psi_f)\rfloor\rfloor} c^i \geq \frac{c(c^{\lfloor\alpha s (1-\psi_f)\rfloor}-1)}{c-1}\\
    & \simeq \frac{c(c^{\alpha s (1-\psi_f)}-1)}{c-1}>\frac{c(c^{\beta s}-1)}{c-1} .
\end{align*}
The last inequality follows from (\ref{throu_bitcoin}). Hence, we conclude that even given some probability of a fork $\psi_f$, an adversary is more likely to succeed attacking Bitcoin than Medium. This implies that, at the same level of security, Medium can tolerate higher mining ratio.

\begin{figure}
    \centering
    \includegraphics[width=0.60\linewidth]{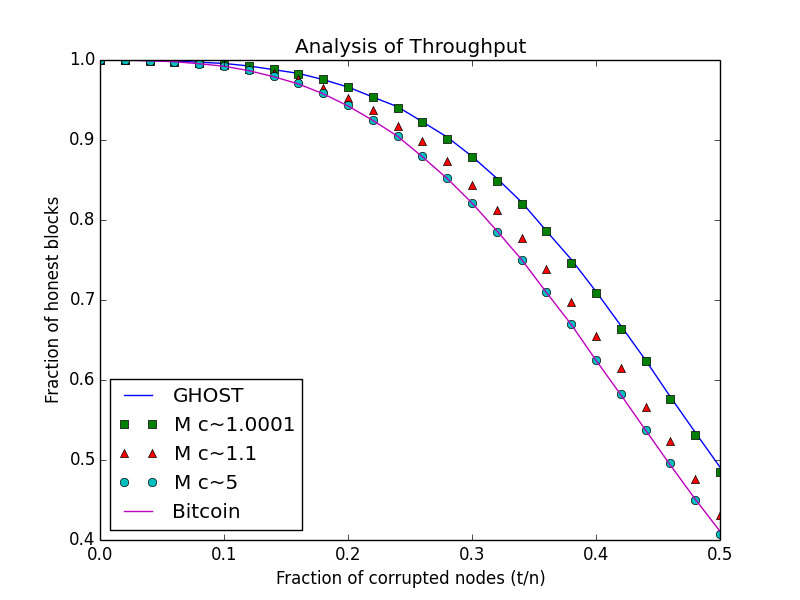} 
    \caption{The fraction of honest blocks in the main chain depending on the number of corrupted parties, with fixed mining ratio such that $npq=1$. The adversary's strategy is to build a heavier secret chain during eight rounds and to release this afterwards. The throughput of Medium (M) is shown for coefficients $c$ of about 5, 1.1, and 1.0001 (the exact values are $\sqrt[10]{10001521}$, $\sqrt[100]{10001521}$, and $\sqrt[100000]{10001521}$, i.e., $n$-th roots of a prime according to Section~\ref{ssec:weightcoefficient}). Throughput is higher with smaller values of~$c$, and the blue line corresponds to GHOST (almost overlapping Medium for $c \approx 5$, and the pink line corresponds to Bitcoin (almost overlapping with Medium for $c \approx 1.0001$). As expected, Medium lies between GHOST and Bitcoin.}
    \label{fig:throughput}
  \end{figure}

Furthermore, we run simulations of the throughput of Medium for different values of $c$ and compare this with Bitcoin and GHOST. The results are shown in Figure~\ref{fig:throughput}. GHOST achieves a higher ratio of honest blocks than both Medium and Bitcoin; however, this has to be contrasted with GHOST's susceptibility to a balance attack, as discussed in the next section.

\section{Analysis of a balance attack}

We first describe the details of the attack that we consider. It is structured as follows, with details shown in Algorithm~\ref{algo:aux}:
\begin{enumerate}
\item The adversary cuts the communication between two sets of parties
  $\CP_1$ and $\CP_2$ with approximately equal hashing power.  This
  partitions the network in two.
\item The honest parties continue running the protocol for $\tau$ rounds,
  but only receive blocks produced within their own partition.  The parties
  build independent subtrees in each partition.
\item During these $\tau$ rounds, the adversary divides its hashing power
  between the partitions. Every block produced by the adversary is added to
  a bank of reserve blocks, $\CB_1$ or $\CB_2$, in the corresponding
  partition.
\item After $\tau$ rounds, the adversary enables communication among all
  parties again and tries to balance the two trees.  This means that it
  releases blocks from the banks (or freshly mined blocks) with the goal of
  preventing that the parties agree on the same main chain across the
  former partitions.  Notice that every block released like this may be
  broadcast selectively, so that it is only received by some honest parties
  initially.  Even if the adversary may not be able to perfectly balance
  the trees with this strategy, it can release blocks to make one tree
  heavier than the other only in the local view of the parties in one
  partition.
\item Once the adversary runs out of blocks in the banks, the attack is
  over and the adversary cannot further balance the trees.  Eventually, the
  honest parties converge on one subtree and on a single chain.
\end{enumerate}

\begin{figure}[!ht]
    \centering
      \includegraphics[width=0.8\linewidth]{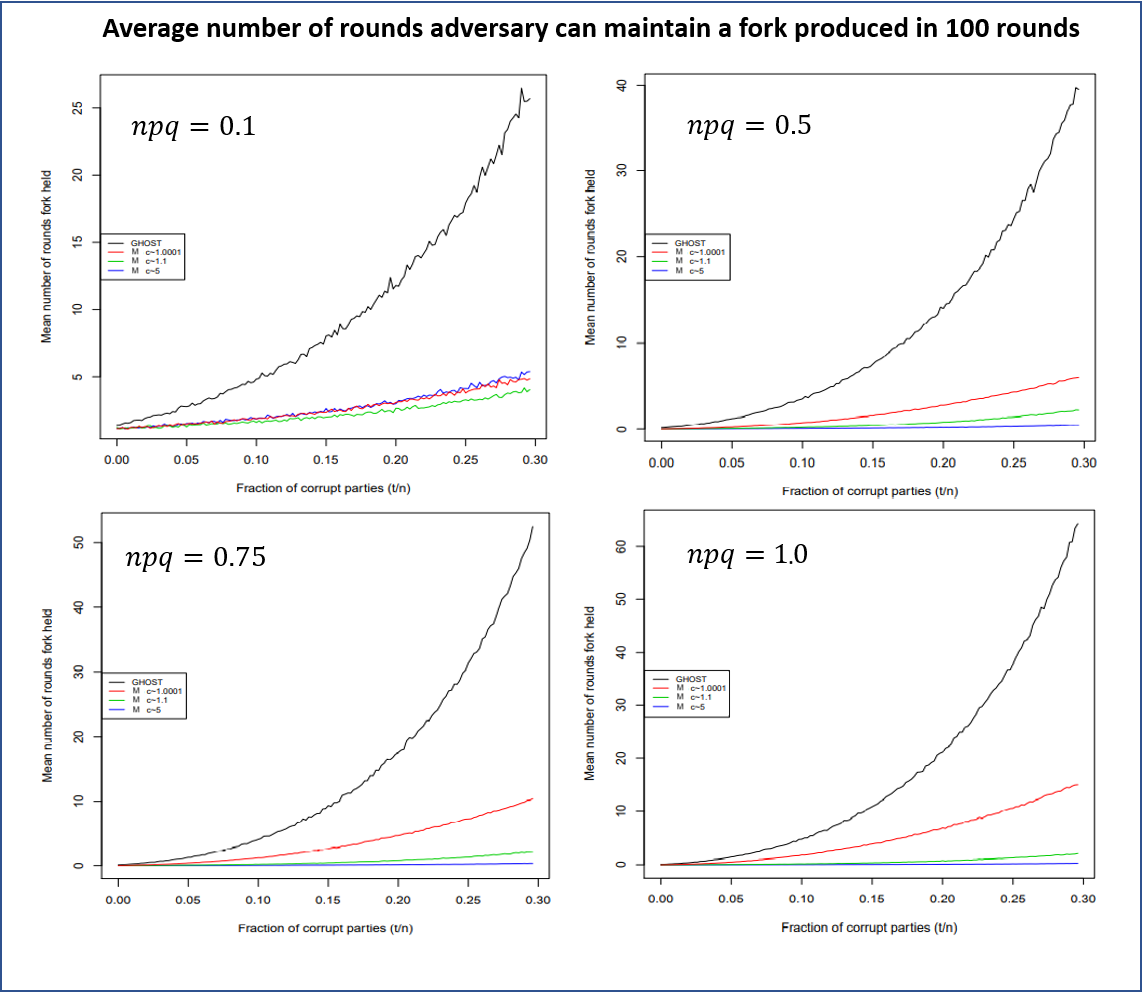}
      \caption{Simulations of how long a fork can be perpetuated by the adversary in the different protocols. A fork is created and maintained for 100 rounds, during which two partitions of the network are isolated from each other. The adversary compensates the weight of the heavier fork greedily, using the smallest number of blocks from its bank. This means that it more likely releases ``heavier'' blocks. The simulation shows four different values of the total mining ratio $pqn$. Again, we show Medium for $c$ of about 5, 1.1, and 1.0001 (as in Figure~\ref{fig:throughput}).
      Notice that with $c$ close to 1, the duration of the fork in Medium is almost an order of magnitude lower than in GHOST and comparable with Bitcoin.}
    \label{fig:BalanceAttackSimulations}
    \end{figure}

Simulations of the resistance of the Medium protocol against this attack are shown in Figure~\ref{fig:BalanceAttackSimulations}. The figure shows for how long the adversary can keep the fork alive and thus prevent the parties from agreeing  after the partition has healed. Since deeper blocks weigh more, the adversarial strategy is to mine as deeply as possible in each partition.  The duration of the fork in Medium can be almost an order of magnitude lower than in GHOST and comparable with Bitcoin.

\begin{algo}
  \small
  \begin{tabbing}
    xxxx\=xxxx\=xxxx\=xxxx\=xxxx\=xxxx\=xxxx\kill
Partition the network in two parts for $\tau$ rounds.\\
Denote the trees in each partition by $T_1$ and $T_2$.\\
Assume $T_1$ and $T_2$ are rooted at blocks $B_1$ and $B_2$.\\
The adversary splits his mining power between partitions\\
Adversary creates banks $\CB_1$ and $\CB_2$. \\
$\ell$ denotes the length of the main chain in subtree $T_i$\\
$n\gets 0$  \` // number of rounds after the first $\tau$\\
\textbf{while} \true \textbf{do}\\
\> \textbf{if} $\exists i, j : [\omega(B_i)>\omega(B_j)]\lor [\omega(B_i)=\omega(B_j) \land \ell(T_i) > \ell(T_j)]$ \textbf{do}\\
\>\> $\Delta\gets \omega(B_i)- \omega(B_j)$\\
\>\> \textbf{if} $\exists \CB' \subseteq \CB_j$ : $[\omega(\CB') \geq \Delta] \land [\ell(T_j \cup \CB') > \ell(T_i \cup \CB')]$ \textbf{do}\\
\> \>  Release subset $\CB'$ of minimal weight to partition $j$ \\
\>\>  \textbf{else} \` //adversary lost\\
\> \>\> \textbf{return} n\\
\> $n \gets n + 1$ \\
\> Honest parties and adversary mine on their respective local view
\end{tabbing}
\caption{Balance attack of $\tau$ rounds}
\label{algo:aux}
\end{algo}

\begin{theorem}
  \label{theo:balance}
Under the assumptions of an $(\varepsilon,\lambda)$-typical execution\footnote{The properties showed in Section~\ref{Security Analysis of Poly GHOST Backbone protocol} may not hold due to the partition of the network. However, the conditions of $(\varepsilon,\lambda)$-typical execution still hold.}, the duration of the balance attack on Medium is bounded by $R \lambda$ rounds, where  where $R$ is the solution of
\[ B_{\tau} = \frac{(c^{R (1 + \varepsilon) \lambda \beta} -1)}{(c^{(1 + \varepsilon) \lambda \beta}-1)}. \]
$B_{\tau}$ is the sum of the number of blocks the adversary has in the banks after $\tau$ rounds, before he has released blocks to balance the fork.

Furthermore, if $\tau \geq \lambda$ then $B_{\tau} < (1 + \varepsilon) p q t \tau$ and the bound can be rewritten as 
\[ R < \frac{\log_c{[(1 + \varepsilon) p q t \tau(c^{(1 + \varepsilon) \lambda \beta}-1)+1]}}{(1 + \varepsilon) \lambda \beta}. \]
\end{theorem}

\begin{proof}
The best case for the adversary is when after each round the two subtrees are equally balanced, in weight and length. It is clear that after each uniquely successful round the tree becomes unbalanced, and thus the adversary is forced to release at least one block to balance the subtrees. For any set of consecutive round of size at least $\lambda$, it holds that $Z(\lambda) < Y(\lambda)$, thus for every $\lambda$ consecutive rounds there is at least one round where the adversary has to broadcast blocks from the bank.

We shall further show that the blocks in an adversary's bank always loose weight over time. We know that the length of the main chain of a subtree $i$ increases by one after a uniquely successful round, to balance the other subtree $j$  the adversary has to release blocks of equivalent or greater weight than the weight of the newly mined block in subtree $i$. If the adversary were to release blocks that balance this block but decrease the length of the main chain in subtree $j$ it has to compensate a greater weight next time an honest party mines on $i$ as the honest parties in $j$ mine at a much lower depth. (Furthermore, if $c$ is chosen so that blocks of a lower depth cannot fully balance blocks of a higher depth, as soon as the adversary releases blocks from the bank that are less deep in the chain it can longer fully balance the two chains and starts to be  force to release blocks in each subsequent round, regardless if an honest mines, meaning an even faster decrease of the bank and the attack failing even earlier.) Thus, we assume the main chains increase in length in any round that the adversary can balance them without using the bank, and the chain thus grow by $Z(\lambda)$ is $\lambda$ rounds. Furthermore, we assume that the adversary can use it's bank in the first uniquely successful round before it has lost weight. After $R \lambda$ rounds the adversary have to release \[\sum_{i=0}^{R-1} c^{ (1 + \varepsilon) \lambda \beta i} = \frac{(c^{R (1 + \varepsilon) \lambda \beta} -1)}{(c^{(1 + \varepsilon) \lambda \beta}-1)} \] blocks from the bank.

If $\tau\geq\lambda$, the new bounds follow from the conditions of $(\varepsilon,\lambda)$-typical execution applied to the first $\tau$ rounds.
\end{proof}

Theorem~\ref{theo:balance} shows that the duration of the attack is bounded. However, the bound may not be tight in almost every execution.

\section{Conclusion}

Medium is a family of protocols that implement a robust transaction
ledger. Medium shares interesting properties with the well-known Bitcoin
and GHOST protocols. More precisely, Medium achieves better throughput
than Bitcoin, but not better than GHOST. However, with a proper choice of
the weight coefficient~$c$, Medium tolerates a balance attack some
orders of magnitude better than GHOST. We conclude that Medium is a
protocol that lies between GHOST and Bitcoin and inherits the good
properties from either side.

Future work may refine the security analysis of Medium, as the properties
established here may not be tight. Alternatively, a Markov-chain based
analysis~\cite{DBLP:conf/ccs/Better} could be used.  Another extension
would be to consider dynamic sets of
parties~\cite{DBLP:journals/iacr/Varying}.

\section*{Acknowledgments}
The authors would like to thanks Jovana Mi\'ci\'c for the support and the interesting discussions.

This work has been funded in part by the Swiss National Science Foundation
(SNSF) under grant agreement Nr\@.~200021\_188443 (Advanced Consensus
Protocols).

\bibliographystyle{acm}
\bibliography{references,dblp_references}

\appendix
\section{Appendix: Protocol Details} 
\label{app:protocols}
We follow the approach of Kiayias and Panagiotakos~\cite{DBLP:conf/latincrypt/KiayiasP17} and use three external functions to describe our protocol, $\op{V}(\cdot), \op{I}(\cdot)$ and $\op{R}(\cdot)$. We call these functions the input validation predicate, the input contribution function and the chain reading function respectively. As in GHOST and Bitcoin, $\op{V}(\cdot)$ controls that the content of a block fulfills specific criteria. We recall that a block is represented in the form $[s, x, i, ctr]$. $\op{V}(\cdot)$  only returns \true if all criteria hold for a block (the contents of a block are given in the $x$ variable). The $\op{I}(\cdot)$ function in its simplest form tells a party what contents should be inserted into the next block to be mined. It receives as input a tuple, $(\var{state},\CM, C, \var{round}, \op{RECEIVE}_i )$, where $\var{state}$ stands for state data, $\CM$ for a set of transactions inputed by the users of the protocol and maintained by the party, $C$ for the main chain, and messages received $\op{RECEIVE}_i$. Finally, the chain reading function $\op{R}(\cdot)$ reads the contents of the main chain $C$. The \op{BROADCAST}() function is the way a party $P_i$ can send a message via the diffusion functionality to all other parties.

  \begin{algo}
    \begin{tabbing}
        xxxx\=xxxx\=xxxx\=xxxx\=xxxx\=xxxx\=xxxx\kill
    $T \leftarrow  \var{genesis}$ \\
    $\var{state} \leftarrow \varepsilon$ \\
    \textbf{for} $\var{round}=1,2,3...$\ \textbf{do} \\
    \>	$[T_{new}, B] \leftarrow \op{update}(T, \text{blocks found in } \op{RECEIVE}_i)$ \\
    \>	$C \leftarrow \op{Medium}(T_{new},\omega_c)$ \` // $\omega_c$ is the global weight function \\
    \>	$[\var{state}, x] \leftarrow \op{I}(\var{state},\CM, C, \var{round}, \op{RECEIVE}_i)$ \\
    \>	$C_{new} \leftarrow \op{POW}(x, i, C)$ \\
    \>	\textbf{if}  $T \ne T_{new}$ \textbf{then} \\
    \> \> 		$\op{BROADCAST}(B)$ \\
    \> \>		$T \leftarrow T_{new}$ \\
    \>	\textbf{if} $C \ne C_{new}$ \textbf{then} \\
    \> \>		$ T \leftarrow \op{update}(T_{new}, \op{head}(C_{new}))$ \\
    \> \> 		$\op{BROADCAST}(\op{head}(C_{new}))$ \\
    \> \textbf{output} $R(C)$ \` // outputs the list of transactions in the chain\\
    \end{tabbing}
    \caption{Medium protocol, as run by honest party $i$.}
  \end{algo}

\begin{algo}
  \begin{tabbing}
      xxxx\=xxxx\=xxxx\=xxxx\=xxxx\=xxxx\=xxxx\kill
	\textbf{function} $\op{PoW}(x, i,  C)$ \\
	\> \textbf{if} $C = \varepsilon$ \textbf{then} \\
	\> \> 	$s \leftarrow 0$ \\
	\> \textbf{else} \\
	\> \>	$[s', x', i',ctr' ] \leftarrow \op{head}(C)$ \\
	\> \> $s \leftarrow \op{H}(ctr', \op{G}(s', x', i'))$ \\
	\> $ctr \leftarrow 1$ \\
	\> $ B \leftarrow \varepsilon$ \\
	\> $h \leftarrow \op{G}(s, x, i)$ \\
	\> \textbf{while} $(ctr \leq q)$ \textbf{do} \\
	\> \> \textbf{if} $(\op{H}(ctr, h) < D)$ \textbf{then} \\
	\> \> \> 	$B \leftarrow [s, x, i, ctr]$ \\
	\> \> \> 	\textbf{break} \\
	\> \>	$ctr \leftarrow ctr + 1$ \\
	\> \textbf{return} $C||B$
	
	\end{tabbing}
  \caption{PoW function, with input $(x, i, C)$, or block content $x$, party $i$ and main chain $C$. This function parameterized by $q$, $D$, and cryptographic hash functions $\op{G}(\cdot)$ and $\op{H}(\cdot)$.}
\end{algo}

\begin{algo}
  \begin{tabbing}
      xxxx\=xxxx\=xxxx\=xxxx\=xxxx\=xxxx\=xxxx\kill
	\textbf{function} $\op{update}(T, B)$ \\
	\> $(B',B^*) \leftarrow (\emptyset,\emptyset)$ \\
	\>	\textbf{for} $[s', x', i', ctr']$ \textbf{in} $B$ \textbf{do}\\
	\> \> 		\textbf{if} $\op{V}(x')$ \textbf{then} \` // input $x$ fulfills validation criteria \\
	\> \>\> 			$B' \leftarrow B' \cup [s', x', i', ctr']$ \\
  \>	\textbf{for} $[s, x, i, ctr]$ \textbf{in} $T$ \textbf{do} \\
	\>\>	\textbf{for} $[s', x', i', ctr']$ \textbf{in} $B'$  \textbf{do}\\
  \> \> \> 	\textbf{if} $s' = \op{H}(ctr, \op{G}(s, x, i))$\\
  \> \> \>$\mbox{}\land\op{H}(ctr', \op{G}(s', x', i')) < D\land ctr^\prime \leq q$ \textbf{then} \\
  \> \> \> \> // $[s', x', i', ctr']$ is valid and extends the tree\\
  \> \> \> \> 		\textbf{insert} $[s', x', i', ctr']$ into $T$\\
  \> \> \> \> $\mbox{}$ as descendent of $[s, x, i, ctr]$\\
	\> \> \> \> 		$B^* \leftarrow B^* \cup [s', x' , i', ctr'] $ \\
	\> \textbf{return} $[T, B^*]$

	\end{tabbing}
        \caption{Tree update function, with input a block tree $T$ and a set of blocks $B$. Further parameters are $q$, $D$, $\op{G}(\cdot)$ and $\op{H}(\cdot)$. }
        \label{alg:update}
\end{algo}

\end{document}
\endinput